\documentclass[times]{aastex7}
\usepackage{amsmath}
\def\lsim{\mathrel{\raise.3ex\hbox{$<$}\mkern-14mu\lower0.6ex\hbox{$\sim$}}}

\begin{document}
\title{Emission and Absorption Features of Magnetically Driven Disk Winds in Black Hole X-Ray Binaries}
\author[0000-0002-0114-5581]{Atsushi Tanimoto}
\affiliation{Department of Science, Kagoshima University, Kagoshima 890-0065, Japan}
\email{atsushi.tanimoto@sci.kagoshima-u.ac.jp}
\author[0000-0001-5709-7606]{Keigo Fukumura}
\affiliation{Department of Physics and Astronomy, James Madison University, Harrisonburg, VA 22807, USA}
\email{fukumukx@jmu.edu}
\author[0000-0002-5701-0811]{Shoji Ogawa}
\affiliation{Institute of Space and Astronautical Science, Japan Aerospace Exploration Agency, Sagamihara, Kanagawa 252-5210, Japan}
\email{sogawa@ac.jaxa.jp}
\author[0000-0003-2670-6936]{Hirokazu Odaka}
\affiliation{Department of Earth and Space Science, Osaka University, Osaka 560-0043, Japan}
\email{odaka@ess.sci.osaka-u.ac.jp}
\author[0000-0002-6562-8654]{Francesco Tombesi}
\affiliation{Physics Department, Tor Vergata University of Rome, Via della Ricerca Scientifica 1, 00133 Rome, Italy}
\affiliation{INAF – Astronomical Observatory of Rome, Via Frascati 33, 00040 Monte Porzio Catone, Italy}
\affiliation{INFN - Rome Tor Vergata, Via della Ricerca Scientifica 1, 00133 Rome, Italy}
\email{francesco.tombesi@roma2.infn.it}
\author[0000-0001-5762-6360]{Marco Laurenti}
\affiliation{Physics Department, Tor Vergata University of Rome, Via della Ricerca Scientifica 1, 00133 Rome, Italy}
\affiliation{INAF – Astronomical Observatory of Rome, Via Frascati 33, 00040 Monte Porzio Catone, Italy}
\affiliation{INFN - Rome Tor Vergata, Via della Ricerca Scientifica 1, 00133 Rome, Italy}
\email{marco.laurenti@roma2.infn.it}
\author[0009-0005-9183-8662]{Pierpaolo Condò}
\affiliation{{Physics Department, Tor Vergata University of Rome, Via della Ricerca Scientifica 1, 00133 Rome, Italy}}
\email{pierpaolo.condo@roma2.infn.it}
\author[0000-0002-1035-8618]{Alfredo Luminari}
\affiliation{INAF – Istituto di Astrofisica e Planetologia Spaziali, Via del Fosso del Caveliere 100, I-00133 Roma, Italy}
\affiliation{INAF – Osservatorio Astronomico di Roma, Via Frascati 33, 00078 Monteporzio, Italy}
\email{alfredo.luminari@inaf.it}

\def\gsim{\mathrel{\raise.5ex\hbox{$>$}\mkern-14mu\lower0.6ex\hbox{$\sim$}}}
\def\lsim{\mathrel{\raise.3ex\hbox{$<$}\mkern-14mu\lower0.6ex\hbox{$\sim$}}}

\begin{abstract}
We investigate accretion disk winds commonly observed in galactic black hole (BH) X-ray binaries (XRB), which manifest as blueshifted absorption features in X-ray spectra. We model these winds as ideal magnetohydrodynamic outflows of hot plasma driven by global magnetic fields threading the accretion disk around the BH. Using Monte Carlo simulations with MONACO, we solve three-dimensional radiative transfer equations to determine the large-scale ionization structure that produces the observed ionic column densities. Focusing on the high/soft state of the BH XRB, where disk emission provides the dominant source of ionizing X-rays, we calculated synthetic spectra showing resonance absorption and scattered emission from ions in various charge states. Our results demonstrate that systems viewed at high polar angles exhibit prominent multi-ion absorption lines with asymmetric profiles, accompanied by P-Cygni-like emission features that partially reproduce the characteristics seen in the observed spectra. This further implies that even a dense disk wind with a high polar angle is unlikely to be saturated due to effective scattering.
\end{abstract}

\section{Introduction}
X-ray binaries (XRBs) are one of the best experimental fields for investigating physical phenomena in strong gravitational fields. \added{In particular, low-mass X-ray binaries (LMXB)} host compact objects such as white dwarfs, neutron stars, or black holes (BHs) accompanied by normal (i.e., Sun-like) ordinary stars. A good proportion of Galactic BH XRBs are known to be transient, undergoing a sequence of unique episodes, and tend to evolve along the well-known q-shaped hysteresis track in the hardness-intensity space. This behavior allows a source to go through a series of distinct accretion states \citep[e.g.,][]{Remillard2006, Belloni2016}, exemplified by high/soft (bright) and low/hard (dim) states, where a dominant spectral component significantly changes despite a diverse pattern of hysteresis varying from burst to burst per source. 

One of the powerful tools for investigating the properties of BH XRBs is X-ray spectroscopic observations. The X-ray spectrum of the BH XRB consists mainly of the following three components. (1) The quasi-thermal spectrum of the accretion disks with a characteristic thermal temperature of $kT \lsim 1~\mathrm{keV}$ \citep{Shakura1973, Mitsuda1984}. (2) The hard X-ray power law component of the photon index $\Gamma \sim 1.5-2.0$, which is generally thought to result from inverse Comptonization in an elusive corona close to a BH \citep{Haardt1991, Remillard2006, Done2007}. (3) A reflection component above $\sim 10~\mathrm{keV}$ is often attributed to the result of the reprocessing of the continuum by the disk through Compton downscattering \citep{Fabian2015}. Reflection spectroscopy is thus found to reshape our understanding of the underlying disk condition (e.g., density, truncation radius, inclination, and BH spin) with high-quality data \citep[e.g.,][]{Garcia2013, Garcia2014}.

It has been well known that LMXB have outflows in the high/soft state. X-ray spectroscopic observations of Chandra High Energy Transmission Grating (HETG) and XMM-Newton Reflection Grating Spectrometer (RGS) have revealed a plethora of ionized absorbers in the form of blueshifted absorption lines, mainly due to the resonant transition of various ions (e.g., O, Ne, Mg, Si, S, and Fe). These X-ray absorbers, most notably detected as H-like ions and He-like ions, have made clear the presence of an outflowing plasma launched from the innermost region of the BH XRBs \citep{Kallman2009, Miller2015a}. The physical properties of the outflow are phenomenologically described by the following three parameters. (1) The hydrogen-equivalent column density along the line of sight $N_{\mathrm{H}}$, (2) the outflow velocity $v_{\mathrm{out}}$, and (3) the ionization parameter $\xi$. Here, the ionization parameter $\xi$ is defined as $\xi \equiv L_{\mathrm{ion}}/(n_{\mathrm{H}} r^2)$ where $L_{\mathrm{ion}}$ is the ionizing luminosity of $1$--$1000 \ \mathrm{Ryd}$, $n_{\mathrm{H}}$ is the hydrogen number density of the wind and $r$ is the distance from the BH. It is commonly known that X-ray winds are preferentially observed in a high/soft state typically exhibiting $\log N_{\mathrm{H}}/\mathrm{cm}^{-2} \sim 21-22$, $\log \xi/\mathrm{erg} \ \mathrm{cm} \ \mathrm{s}^{-1} \sim 3-5$ and $v \lsim 0.01 c$ where $c$ is the speed of light, while leaving little or no wind signatures during the low/hard state \citep{Esin1997, Ponti2012, Ratheesh2021, Fukumura2021} in the evolution of canonical hysteresis (i.e., wind dichotomy \citealt{Neilsen2009}). However, the driving mechanism of the BH XRB outflow remains unclear.

\added{One of the leading candidates for the driving mechanism of the outflow is the magnetic process. In this scenario, a poloidal magnetic field is impinged on the accretion disk, and the disk wind is driven primarily by magnetic centrifugal forces. Previous studies have performed one-dimensional photoionization equilibrium calculations based on self-similar solutions of magnetically driven disk winds and investigated the energies and equivalent widths of the observed absorption lines \citep[e.g.,][]{Blandford1982, Contopoulos1994, Ferreira1997, Fukumura2010, Fukumura2017, Fukumura2021, Fukumura2022}. However, self-similar solutions of magnetically driven disk winds possess a high density that is optically thick to Compton scattering in the equatorial direction. Therefore, to evaluate the central energy and equivalent width of the observed absorption lines, it is essential to incorporate the effects of scattered components through three-dimensional X-ray radiative transfer calculations}.

\added{In this study, we performed three-dimensional X-ray radiative transfer calculations based on self-similar solutions of magnetically driven disk winds with the Monte Carlo Simulations for Astrophysics and Cosmology (MONACO: \citealt{Odaka2016}) code. The remainder of this paper is organized as follows. \hyperref[Section0200]{Section~2} presents the our method. \hyperref[Section0201]{Section~2.1} presents the self-similar solutions of magnetically driven disk winds. \hyperref[Section0202]{Section~2.2} describes the photoionization equilibrium calculations. \hyperref[Section0203]{Section~2.3} presents the three-dimensional X-ray radiative transfer calculations. \hyperref[Section0300]{Section~3} describes the results of these calculations. \hyperref[Section0301]{Section~3.1} shows the distributions of the hydrogen number density and the radial velocity based on the wind model. \hyperref[Section0302]{Section~3.2} presents the distribution of the turbulent velocity, the temperature, and the ionization parameter. \hyperref[Section0303]{Section~3.3} shows the transmitted component and the scattered component obtained from the X-ray radiative transfer calculation. \hyperref[Section0401]{Section~4.1} discusses the origin of the absorption lines. \hyperref[Section0402]{Section~4.2} presents the origin of the neutral iron fluorescent lines. \hyperref[Section0403]{Section~4.3} discusses the effect of scattering. \hyperref[Section0404]{Section~4.4} presents the P-Cygni features. \hyperref[Section0500]{Section~5} describes our conclusions}.\clearpage
\begin{figure*}\label{Figure0001}
\gridline{\fig{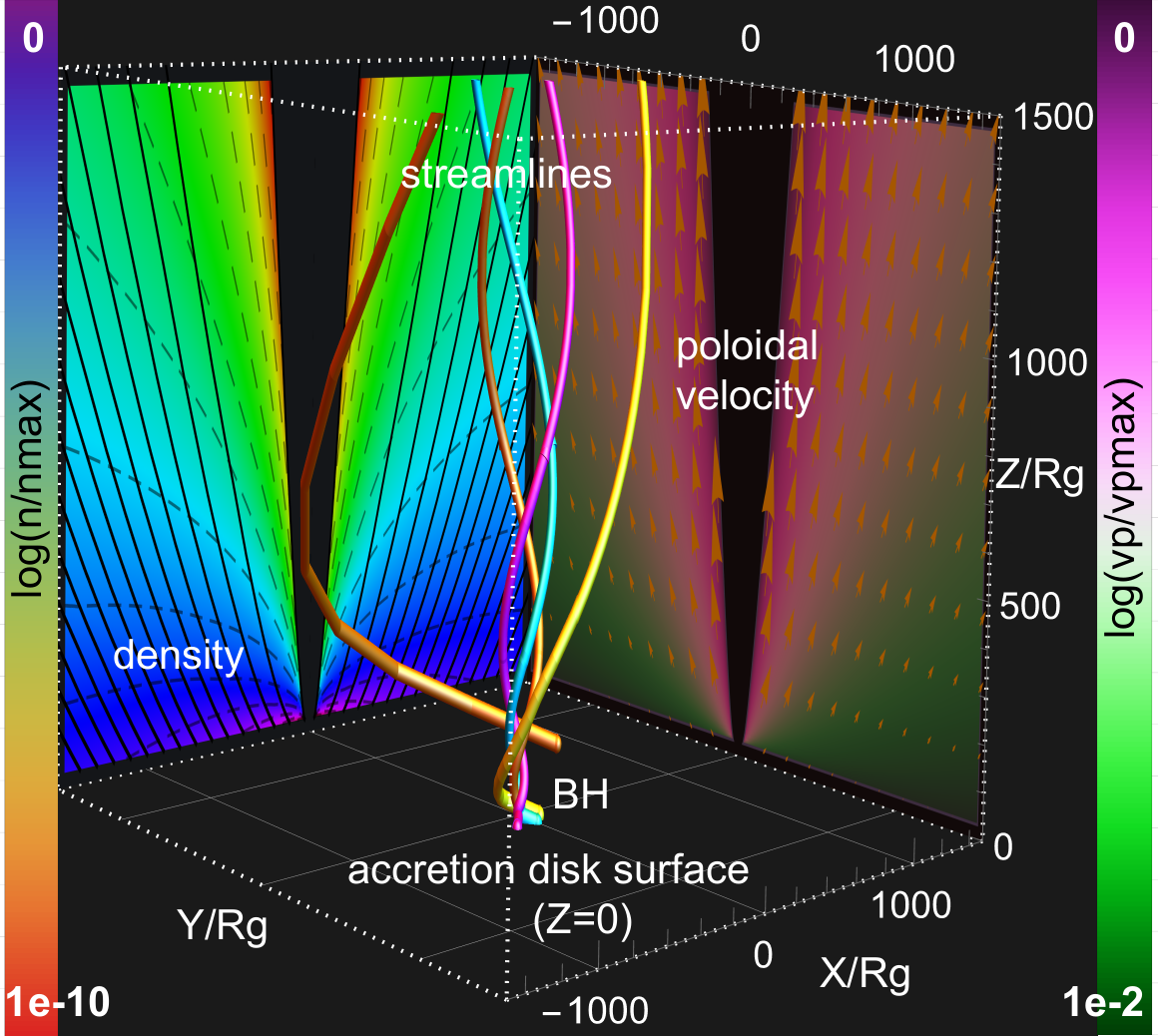}{0.5\textwidth}{}}
\caption{Three-dimensional rendering of five individual streamlines launched magnetically from different locations in the innermost region where $R_{\mathrm{g}}$ is gravitational radius. On X-Z plane, we show the poloidal distribution of the normalized hydrogen number density (color-coded), $(\log n- \log n_{\mathrm{min}})/(\log n_{\mathrm{max}}- \log n_{\mathrm{min}})$, density contours (dashed) and the poloidal projection of magnetic field lines (thick solid). Y-Z plane shows the distribution of the normalized poloidal wind velocity (color-coded), $(\log vp - \log vp_{\mathrm{min}})/(\log vp_{\mathrm{max}}- \log vp_{\mathrm{min}})$, with arrows.}
\end{figure*}
\begin{figure*}\label{Figure0002}
\gridline{\fig{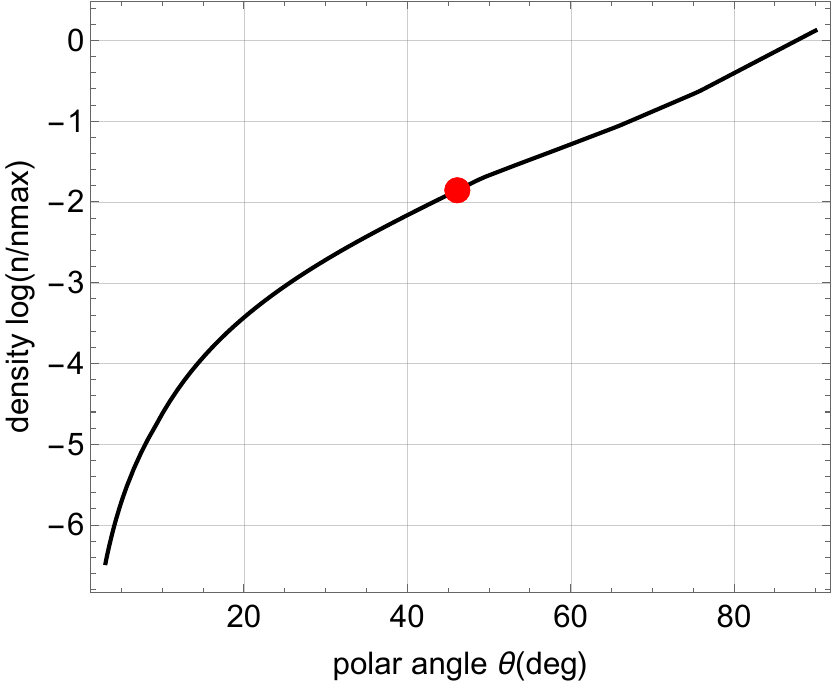}{0.5\textwidth}{(a)}\fig{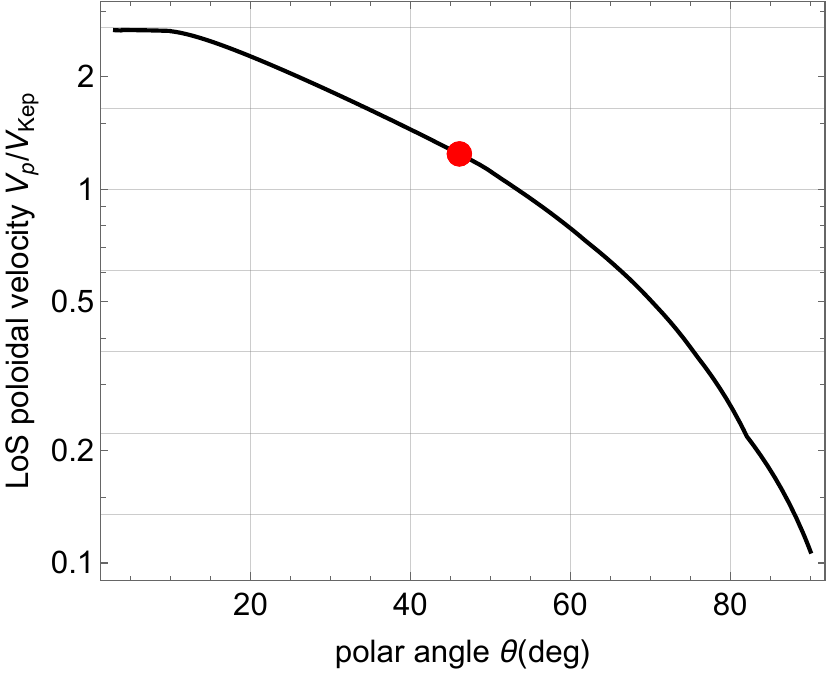}{0.5\textwidth}{(b)}}
\caption{\added{(a) Angular density profile $f(\theta)$ (normalized to unity) given by Equation 1 and (b) angular poloidal velocity profile $g_p(\theta)$ (in units of Keplerian velocity at the base of the wind) given by Equation 2. Red dot denotes the position of the Alfven point.}}
\end{figure*}\clearpage
\section{Methods}\label{Section0200}
This section describes the methodology. \hyperref[Section0201]{Section~2.1} presents the self-similar solutions of magnetically driven disk winds. \hyperref[Section0202]{Section~2.2} describes the photoionization equilibrium calculations. \hyperref[Section0203]{Section~2.3} presents the three-dimensional X-ray radiative transfer calculations.
\subsection{Self-similar Magnetically-driven Disk Winds}\label{Section0201}
Our calculations are based on self-similar magnetically driven disk winds \citep[e.g.,][]{Blandford1982, Contopoulos1994, Ferreira1997, Fukumura2010, Fukumura2017, Fukumura2021, Fukumura2022, Chakravorty2016}. The essential part of the formalism is the radial dependence of the density $n(r,\theta)$ and the velocity components $v_i(r,\theta)$ where $i=r,\theta$ and $\phi$. These are given in spherical coordinates as
\begin{align}
\label{Equation0001}
n_{\mathrm{H}}(r,\theta)     & = \left(\frac{n}{10^{17} \ \mathrm{cm}^{-3}}\right) \left(\frac{r}{R_{\mathrm{in}}} \right)^{-p} f(\theta) \ 10^{17} \ \mathrm{cm}^{-3}\\
v_i(r,\theta)   & = v_{\mathrm{K}}(R_{\mathrm{in}}) \left(\frac{r}{R_{\mathrm{in}}}\right)^{-\frac{1}{2}} g_i(\theta)
\end{align}
where $n$ is the wind density at the footpoint of a streamline launched from the innermost radius at $r = R_{\mathrm{in}}$, which is typically identified with the innermost stable circular orbit (ISCO), $p$ is the radial dependence of the wind density, $f(\theta)$ is the angular dependence of the density, $v_{\mathrm{K}}(R_{\mathrm{in}})$ is the Keplerian velocity at $R_{\mathrm{in}}$, and $g_i(\theta)$ is the angular dependence of the velocity \citep[e.g.,][]{Fukumura2021}. \added{Here, since we assumed axisymmetric solution, the density $n(r,\theta)$ and the velocity $v_i(r,\theta)$ do not depend on the azimuthal angle $\phi$. We also assumed that the line of sight obeys the Keplerian law in our formalism}. 

We calculated the hydrogen number density and the three-dimensional velocity based on Equations \hyperref[Equation0001]{1} and \hyperref[Equation0001]{2}. \hyperref[Figure0001]{Figure~1} shows the three-dimensional rendering of individual streamlines for a fiducial MHD-driven wind solution. We assumed a BH mass of $10~M_{\sun}$. To reduce the degrees of freedom in the wind model, we fixed the normalization of the density of the winds in $n = 1.4 \times 10^{18}~\mathrm{cm}^{-3}$ and the slope of the density $p = 1.2$ (often derived observationally by wind absorption spectroscopy in BH XRBs and active galactic nuclei (AGNs); \citealt{Fukumura2017, Fukumura2021, Ratheesh2021}) in the present exercises.

\added{The angular dependencies $f(\theta)$ and $g_i(\theta)$ are coupled to the geometry of the magnetic field. We determine these angular dependencies by numerically solving the Grad-Shafranov equation combined with the ideal magnetohydrodynamic (MHD) equations (e.g., \citealt{Contopoulos1994} for the original formulation and \citealt{Fukumura2014, Fukumura2017, Fukumura2021} for a more detailed description of the wind morphology). \hyperref[Figure0002]{Figure~2} shows that (a) the angular dependence of the normalized wind density $f(\theta)$ and (b) the angular dependence of the poloidal velocity $g_p(\theta)$ in units of Keplerian velocity at the base of the wind along the innermost field line. \hyperref[Figure0002]{Figure~2(a)} indicates that the wind density varies with polar angle over roughly six decades in magnitude between the equator and the pole, showing that the gas is almost optically thick at the base, while optically thin near the funnel region. \hyperref[Figure0002]{Figure~2(b)} shows that the wind in the poloidal plane is found to monotonically accelerate from sub-Alfvenic regime near the base to super-Alfvenic regime at high latitude along the field line as a consequence of magnetic driving via Lorentz force $\mathbf{J} \times \mathbf{B}$. Note that our line-of-sight intercepts a range of velocities, which necessarily imprints a characteristic blueshift of asymmetry in the line profile, possibly as a telltale sign for MHD winds}.

We divided the radial direction into logarithmic divisions of $40$ that range from $10^{10}$ to $10^{14} \ \mathrm{cm}$, but focused on the relevant radial domain up to $10^{12} \ \mathrm{cm}$. These values correspond to $3.4 \times 10^{4} R_g$ to $3.4 \times 10^{8} R_g$, $R_g$ is defined as $R_g = 2GM_{\mathrm{BH}}/c^2$ where $G$ is the gravitational constant, $M_{\mathrm{BH}}$ is the mass of BH, and $c$ is the speed of light. We also divided the angular direction into seven divisions $\theta = 30\degr, \ 40\degr, \ 50\degr, \ 60\degr, \ 70\degr, \ 80\degr, \ \mathrm{and} \ 89\degr$. In this framework, no winds are launched within the innermost magnetic field line anchored to the innermost launching radius (set to the ISCO radius for a Schwarzschild BH); no winds exist in the polar region. Thus, we consider an angular domain with a polar angle between $25\degr$ and $90\degr$.\clearpage
\subsection{Photoionized Equilibrium Calculations with XSTAR}\label{Section0202}
\added{We performed photoionization equilibrium calculations based on the density distribution obtained from our wind solutions with XSTAR \citep{Kallman2001}. The XSTAR is a code for calculating the physical conditions and spectra of ionized and nearly-neutral gases. XSTAR assumes a plane-parallel slab geometry during its photoionization calculations. In photoionization equilibrium calculations, XSTAR considers ionization processes such as photoionization, collisional ionization, and autoionization, and recombination processes such as radiative recombination, three-body recombination, and dielectric recombination. To determine the temperature, XSTAR solves the balance between heating via photoelectric heating and Compton heating and cooling via radiative deexcitation, radiative recombination, and bremsstrahlung. Since XSTAR is a one-zone photoionization equilibrium calculation code, we performed XSTAR calculations for each grid and calculated the photoionization equilibrium from the inside to the outside}.

\begin{figure*}\label{Figure0003}
\gridline{\fig{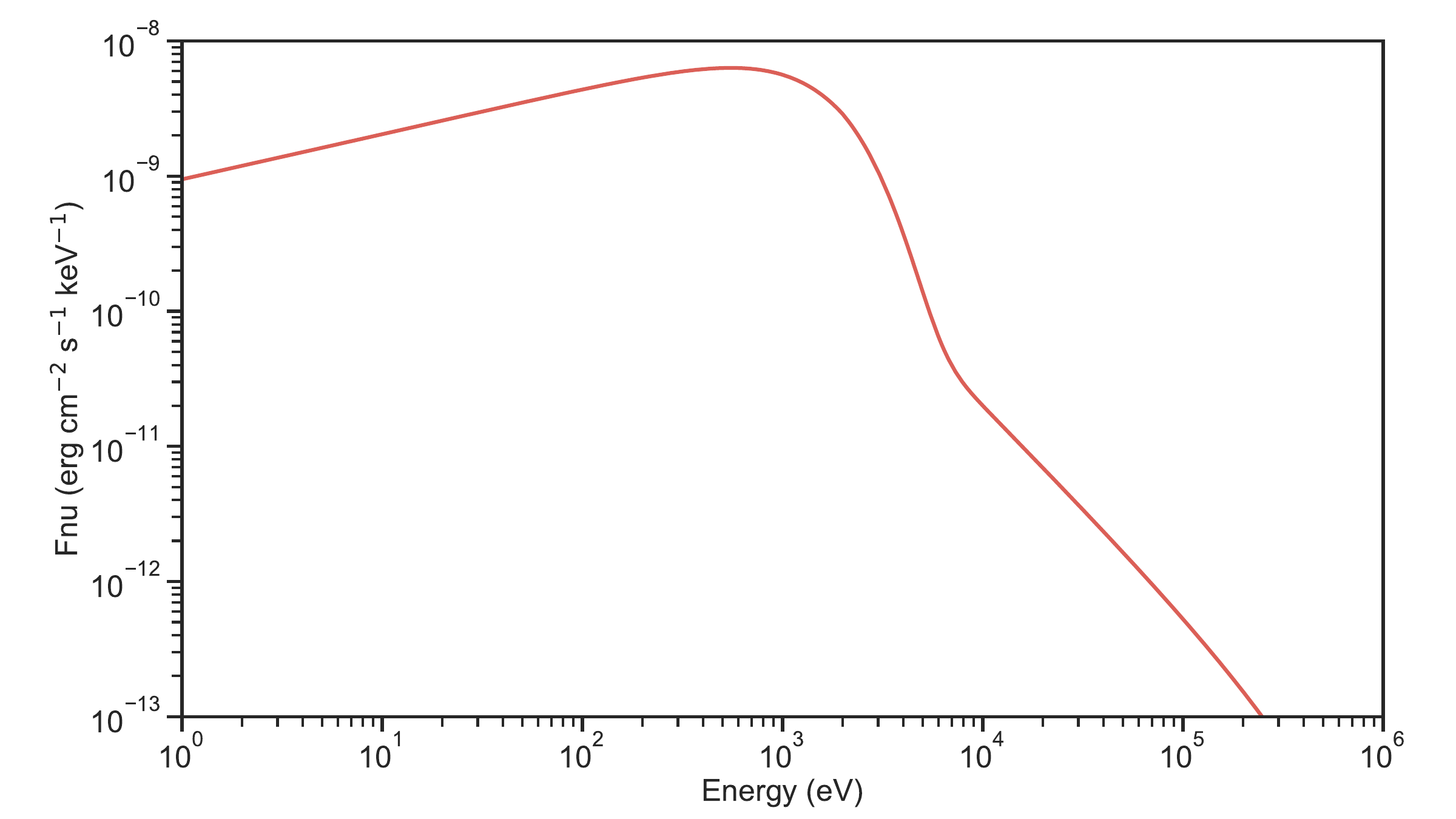}{0.5\textwidth}{}}
\caption{The shape of the adopted X-ray spectrum of a fiducial high/soft state \citep[e.g.,][for H1743--322]{Shidatsu2019}.}
\end{figure*}

\added{The following five parameters are required to perform XSTAR calculations. (1) inner radius, (2) outer radius, (3) hydrogen number density in the region, (4) shape of the X-ray spectrum, and (5) X-ray luminosity at $1$--$1000 \ \mathrm{Ry}$. (1)--(2) We generate a total of $280$ grids, with the $10^{10}$ to $10^{14}~\mathrm{cm}$ region divided $40$ in the radial direction on logarithmic scale and the $25\degr$ to $90\degr$ region divided seven in the angular direction. (3) We assume the hydrogen number density based on the wind solutions. (4)--(5) We adopt the X-ray spectrum of H1743--322 in the high/soft state observed by Chandra HETG \citep{Shidatsu2019} as a fiducial case. This is because H1743--322 is a well-studied BH XRB \citep{Homan2005, McClintock2009, Ingram2016, Shidatsu2019, Tomaru2019}, and X-Ray Imaging and Spectroscopy Mission (XRISM; \citealt{Tashiro2025}) is scheduled to observe this object}. The model is expressed in XSPEC terminology as
\begin{equation}
\mathtt{TBabs} \times \mathtt{simpl} \times \mathtt{diskbb} \ ,
\end{equation}
where $\mathtt{TBabs}$ represents the absorption by the interstellar medium \citep{Wilms2000} of the hydrogen column density of $N_H = 1.60 \times 10^{22} \ \mathrm{cm}^{-2}$ \citep{Capitanio2009}. $\mathtt{simpl}$ corresponds to the Comptonization model which convolves a fraction of an input spectrum into a power law with the photon index $\Gamma$ and the fraction of the total input X-ray flux scattered $F_{\mathrm{scat}}$ \citep{Steiner2009}. $\mathtt{diskbb}$ represents the multicolor disk blackbody radiation \citep{Mitsuda1984} characterized by the temperature of the innermost disk $T_{\mathrm{in}}$. In the high/soft state of H1743--322, the best fitting parameters have been derived as follows; $\Gamma = 2.27 \pm 0.06$, $F_{\mathrm{scat}} = 2.40_{-0.10}^{+0.20} \times 10^{-2}$, and $T_{\mathrm{in}} = 1.221 \pm 0.002 \ \mathrm{keV}$. The corresponding X-ray luminosity ($1$--$1000 \ \mathrm{Ry}$) is $L_{\mathrm{X}} = 3.50 \times 10^{38} \ \mathrm{erg} \ \mathrm{s}^{-1}$ \citep[e.g.,][]{Shidatsu2019}. \hyperref[Figure0003]{Figure~3} shows the X-ray spectrum of H1743--322 during the high/soft state where the $\mathtt{diskbb}$ component is dominant.\clearpage
\subsection{X-Ray Radiative Transfer Calculations}\label{Section0203}
To perform X-ray radiative transfer calculations, we used the MONACO code version 1.8.2. The MONACO code uses the Geant4 code to track photons in complex geometric structures \citep{Agostinelli2003, Allison2006, Allison2016}. Although Geant4 implements physical processes, MONACO uses optimized physical processes for astrophysics. MONACO incorporates three sets of physical processes: (1) X-ray reflection from neutral matter \citep{Odaka2011, Tanimoto2019, Tanimoto2022, Tanimoto2023}, (2) Comptonization in a hot flow \citep{Odaka2013, Odaka2014}, and (3) photon interactions in a photoionized plasma \citep{Watanabe2006, Hagino2015, Hagino2016, Tomaru2018, Tomaru2020, Tomaru2023, Mizumoto2021, Tanimoto2025}.

\added{In this study, we considered (3) photon interactions in a photoionized plasma. In this case, photons emitted randomly from the origin interact with surrounding matter through processes such as photoexcitation, photoionization, and Compton scattering. Since MONACO assumes photoionization equilibrium, radiative deexcitation occurs alongside photoexcitation, and radiative recombination occurs alongside photoionization. To calculate these processes, an atomic database is required. MONACO utilizes ion energy levels, photoexcitation occurrence probabilities, photoionization cross sections, and other parameters generated by the Flexible atomic code \citep{Gu2008}}.

We also took into account Doppler effects due to the motion of the wind. The absorption lines are broadened because of thermal and turbulent motion. Since the line of sight intercepts a gradient of dynamical wind velocity, shear velocity also comes into play for broadening. We defined the turbulent velocity as follows.
\begin{align}
v_{\mathrm{turb}}   & = \max (v_{\mathrm{sound}}, v_{\mathrm{shear}})                                                       \\
v_{\mathrm{sound}}  & = 10 \times \left(\frac{T}{10^4 \ \mathrm{K}}\right)^{\frac{1}{2}} \ \mathrm{km} \ \mathrm{s}^{-1}
\end{align}
Here, $v_{\mathrm{sound}}$ is the speed of sound and $v_{\mathrm{shear}}$ is the dynamical shear velocity. The speed of sound is expressed as in Equation~5 using the temperature obtained from the photoionization equilibrium calculations. For example, in the photoionization equilibrium where \ion{Fe}{25} and \ion{Fe}{26} exist, the plasma temperature is approximately $10^6 \ \mathrm{K}$, which corresponds to the speed of sound of approximately $100 \ \mathrm{km} \ \mathrm{s}^{-1}$. The wind medium in our MHD formalism is generally subject to dynamical shear motion along a line of sight manifested as a velocity gradient of roughly $\sim 10-20\%$ of the line of sight velocity itself, often resulting in a relatively high velocity up to thousands of $\mathrm{km} \ \mathrm{s}^{-1}$ at smaller distances near the equatorial disk around BH, while smaller than $\sim 100 \ \mathrm{km} \ \mathrm{s}^{-1}$ at larger distances.

We perform X-ray radiative transfer calculations using a total of $12,800$ grids. These included $40$ grids for the radial range from $10^{10}$ to $10^{14} \ \mathrm{cm}$, $8$ grids for the angular range from $0$ to $\pi/2$, $40$ grids for the azimuthal angle range from $0$ to $2\pi$. Each grid has physical quantities such as the hydrogen number density and the three-dimensional velocities obtained from the wind model, as well as the ion fractions derived from the photoionization equilibrium calculations. We considered 13 abundant elements such as H, He, C, N, O, Ne, Mg, Si, S, Ar, Ca, Fe, and Ni. Since we are mainly interested in absorption and emission lines ranging from $6$ to $7 \ \mathrm{keV}$, we took into account H-like and He-like ions for carbon, nitrogen, oxygen, neon, magnesium, silicon, sulfur, and calcium, and from H-like to Ne-like ions for iron and nickel. We generated a total of four billion photons from a source at the origin, with energies ranging from $2$ to $10 \ \mathrm{keV}$.

\clearpage
\begin{figure*}\label{Figure0004}
\gridline{\fig{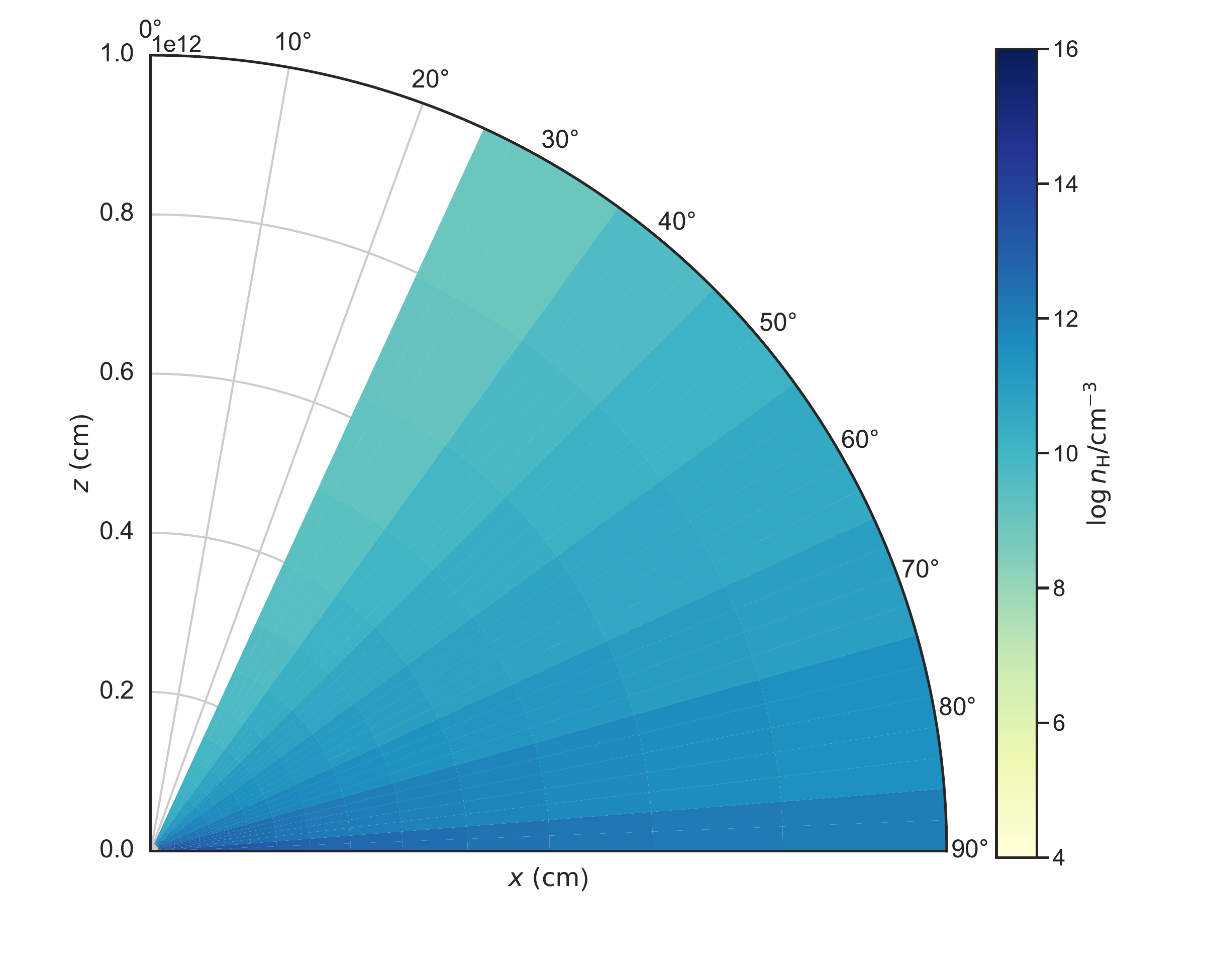}{0.5\textwidth}{(a)}\fig{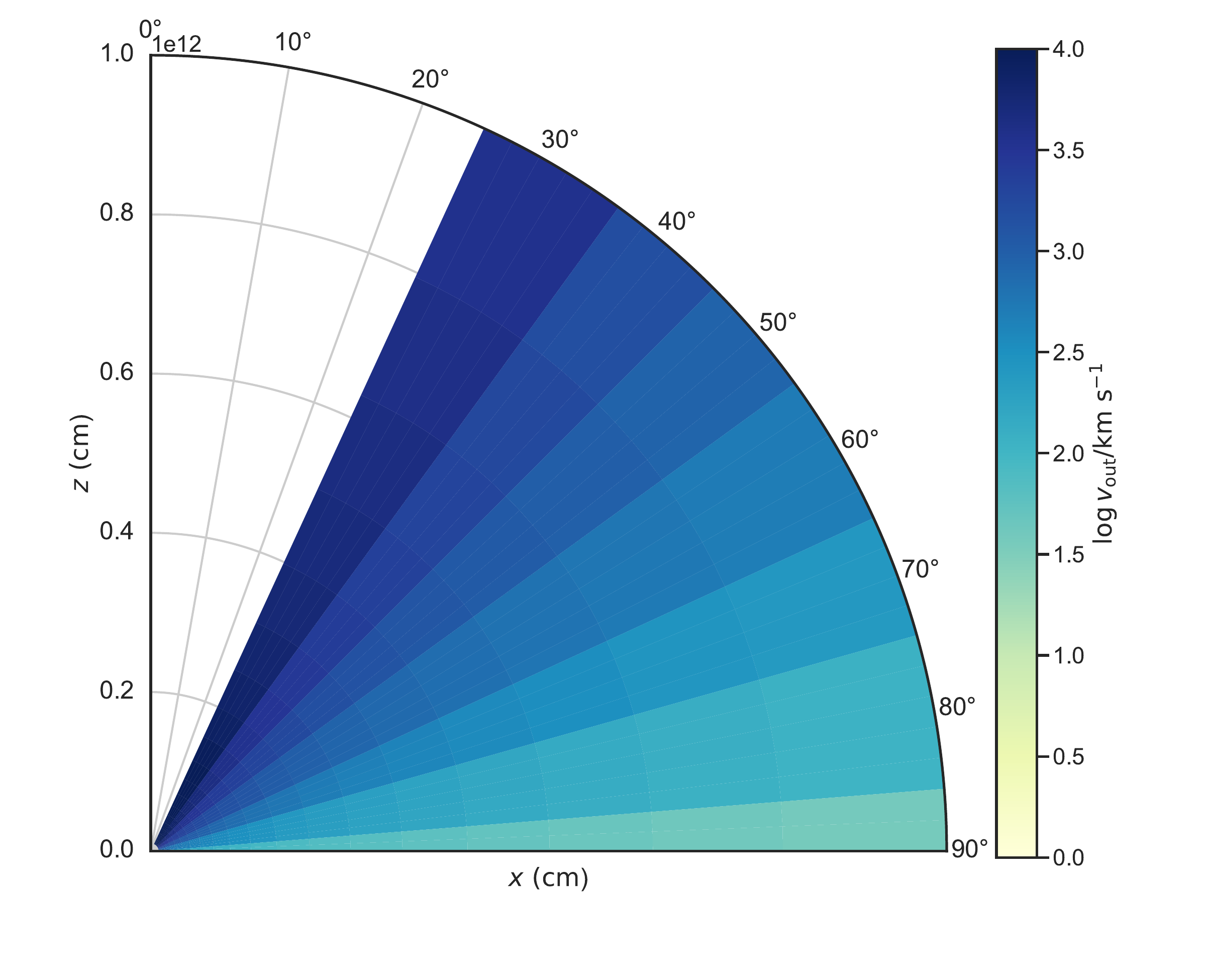}{0.5\textwidth}{(b)}}
\gridline{\fig{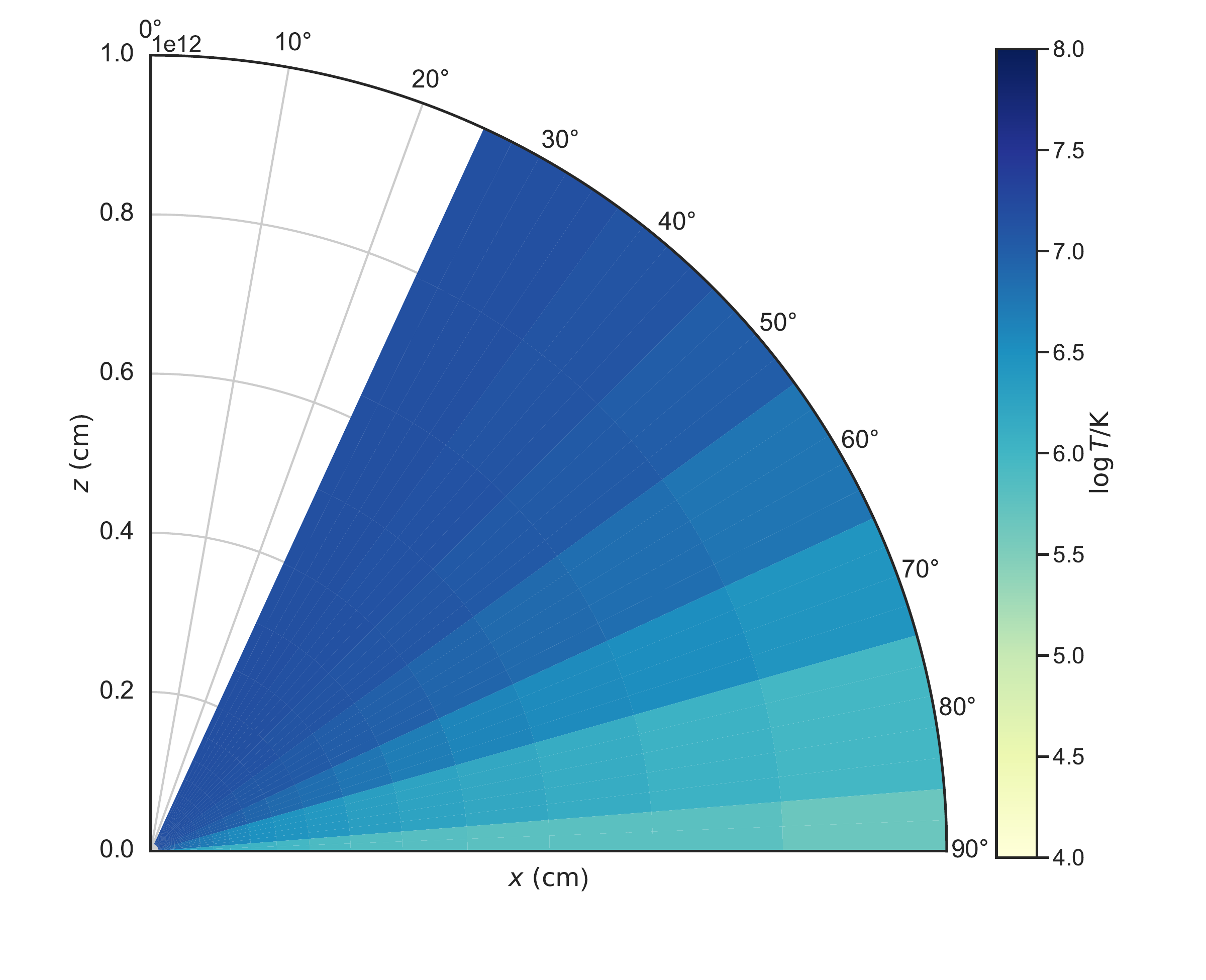}{0.5\textwidth}{(c)}\fig{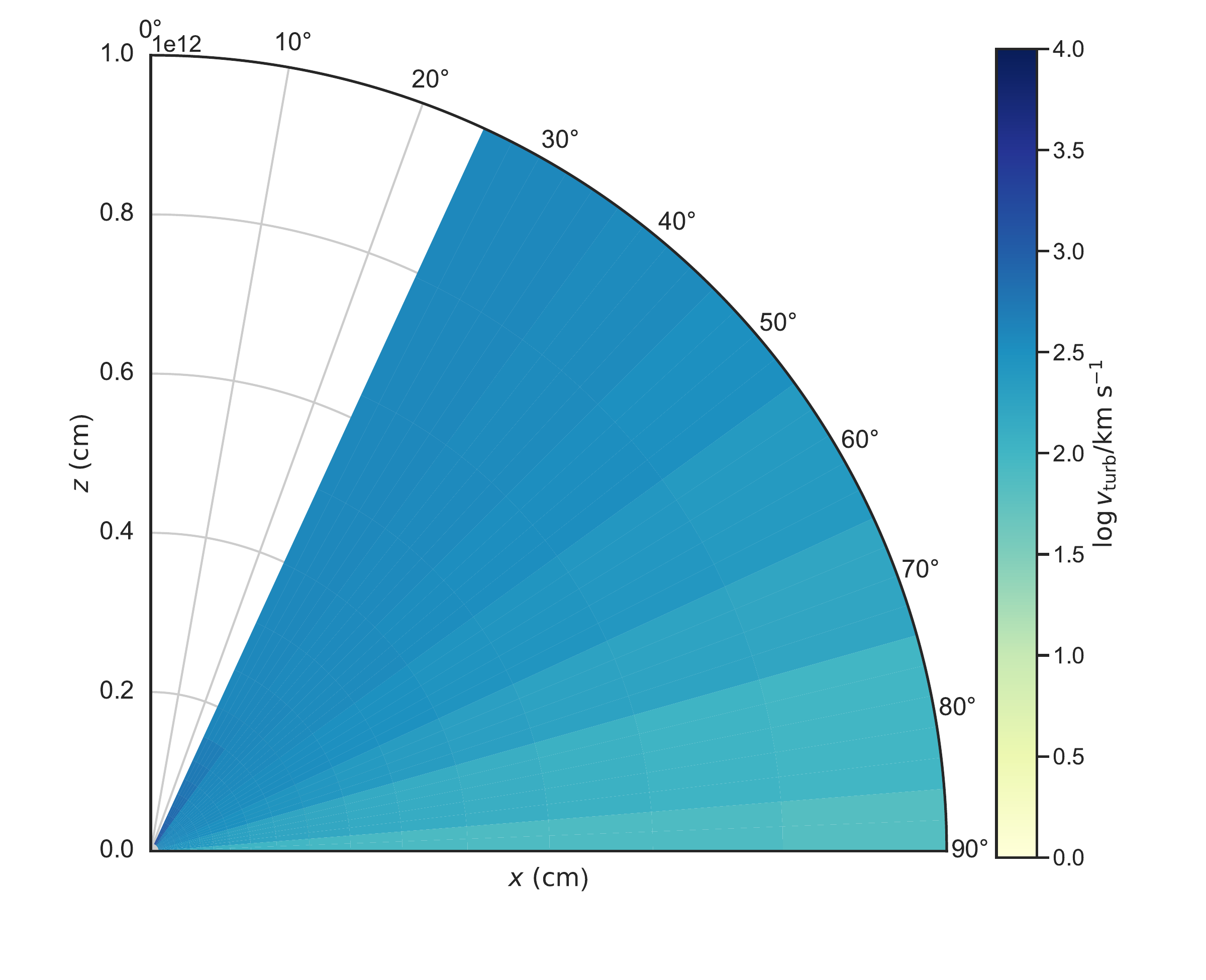}{0.5\textwidth}{(d)}}
\caption{(a) The distribution of the hydrogen number density $\log n_{\mathrm{H}}/\mathrm{cm}^{-3}$. (b) The distribution of the radial velocity $\log v_{\mathrm{out}}/\mathrm{km} \ \mathrm{s}^{-1}$. (c) The distribution of the temperature $\log T/\mathrm{K}$. (d) The distribution of the turbulent velocity $\log v_{\mathrm{turb}}/\mathrm{km} \ \mathrm{s}^{-1}$.}
\end{figure*}
\begin{figure*}\label{Figure0005}
\gridline{\fig{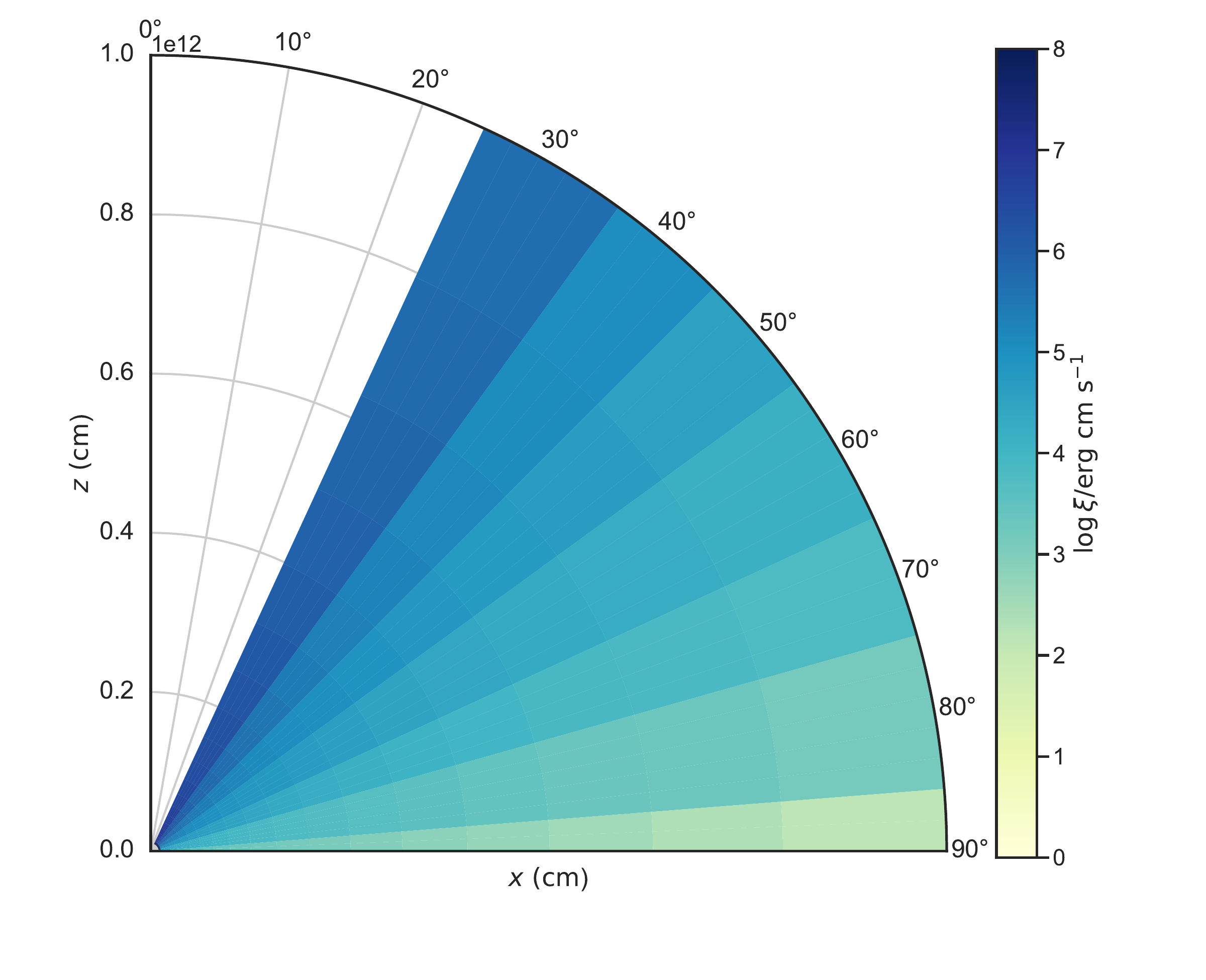}{0.5\textwidth}{}}
\caption{The distribution of the ionization parameter $\log \xi/\mathrm{erg} \ \mathrm{cm} \ \mathrm{s}^{-1}$.}
\end{figure*}\clearpage
\begin{figure*}\label{Figure0006}
\gridline{\fig{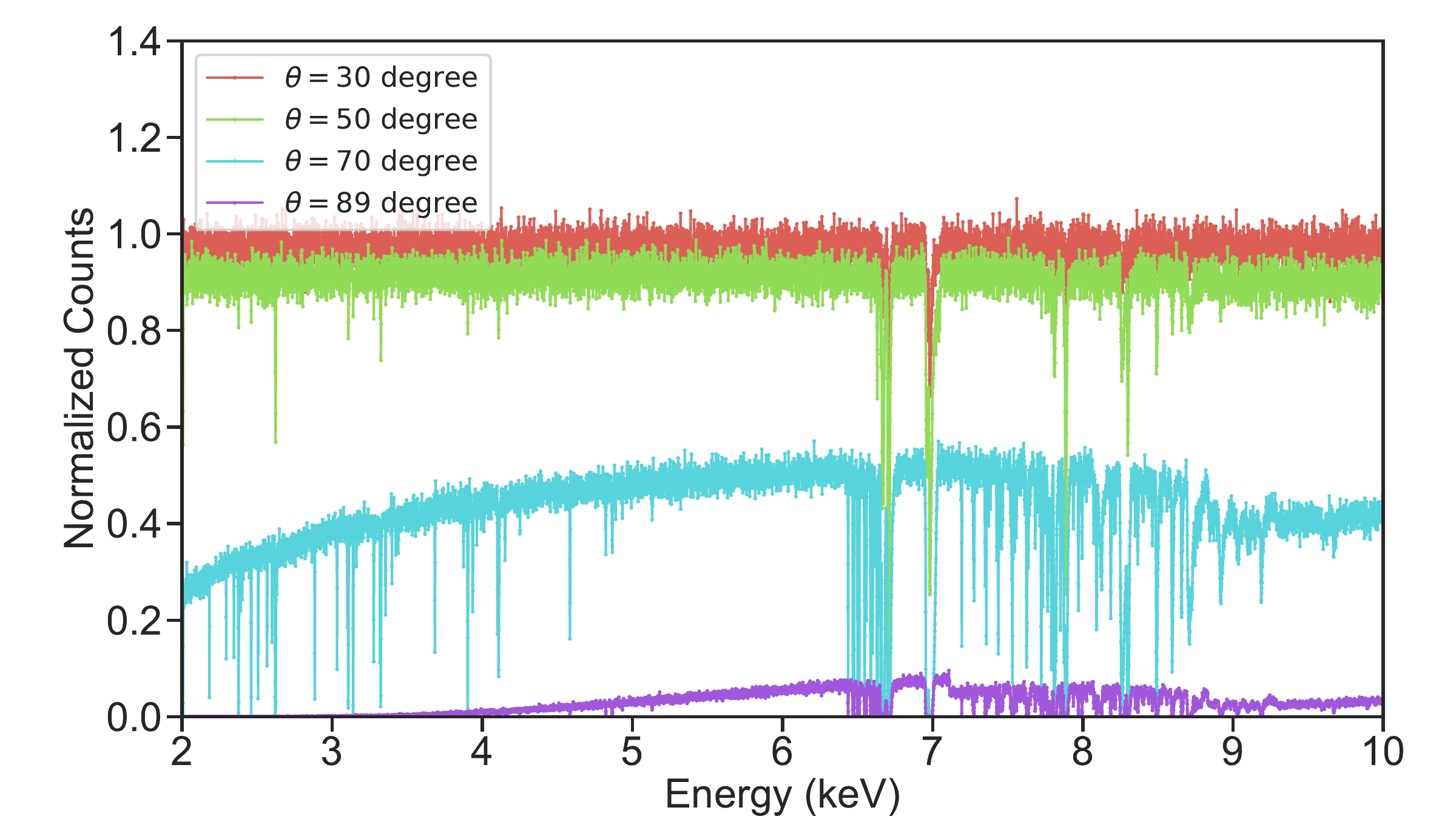}{0.5\textwidth}{(a)}\fig{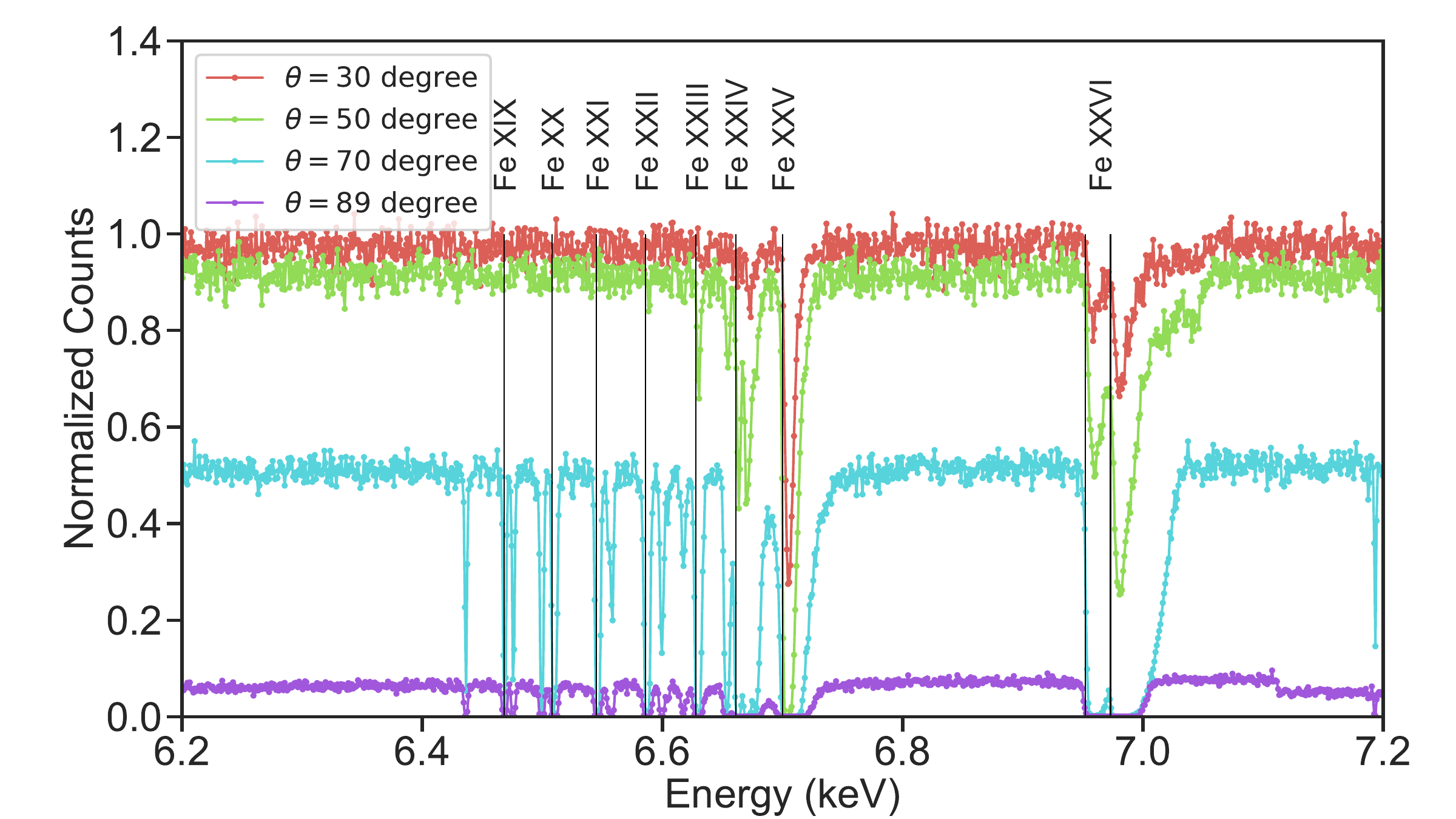}{0.5\textwidth}{(b)}}
\caption{(a) The transmitted component of the X-ray spectrum at $2$--$10 \ \mathrm{keV}$ in the rest frame. The red, green, light blue, and purple line show the X-ray spectrum at the polar angle $\theta = 30\degr$, $50\degr$, $70\degr$, and $89\degr$, respectively. (b) The transmitted component of the X-ray spectrum at $6.2$--$7.2 \ \mathrm{keV}$. The black lines correspond to the energy of the absorption lines in the rest frame due to \ion{Fe}{19} ($6.468 \ \mathrm{keV}$), \ion{Fe}{20} ($6.508 \ \mathrm{keV}$), \ion{Fe}{21} ($6.545 \ \mathrm{keV}$), \ion{Fe}{22} ($6.586 \ \mathrm{keV}$), \ion{Fe}{23} ($6.628 \ \mathrm{keV}$), \ion{Fe}{24} ($6.661 \ \mathrm{keV}$), \ion{Fe}{25} ($6.700 \ \mathrm{keV}$), \ion{Fe}{26} Ly$\alpha$2 ($6.952 \ \mathrm{keV}$), and \ion{Fe}{26} Ly$\alpha$1 ($6.973 \ \mathrm{keV}$). Energies of absorption lines are based on the Flexible Atomic Code \citep{Gu2008}}
\end{figure*}
\begin{figure*}\label{Figure0007}
\gridline{\fig{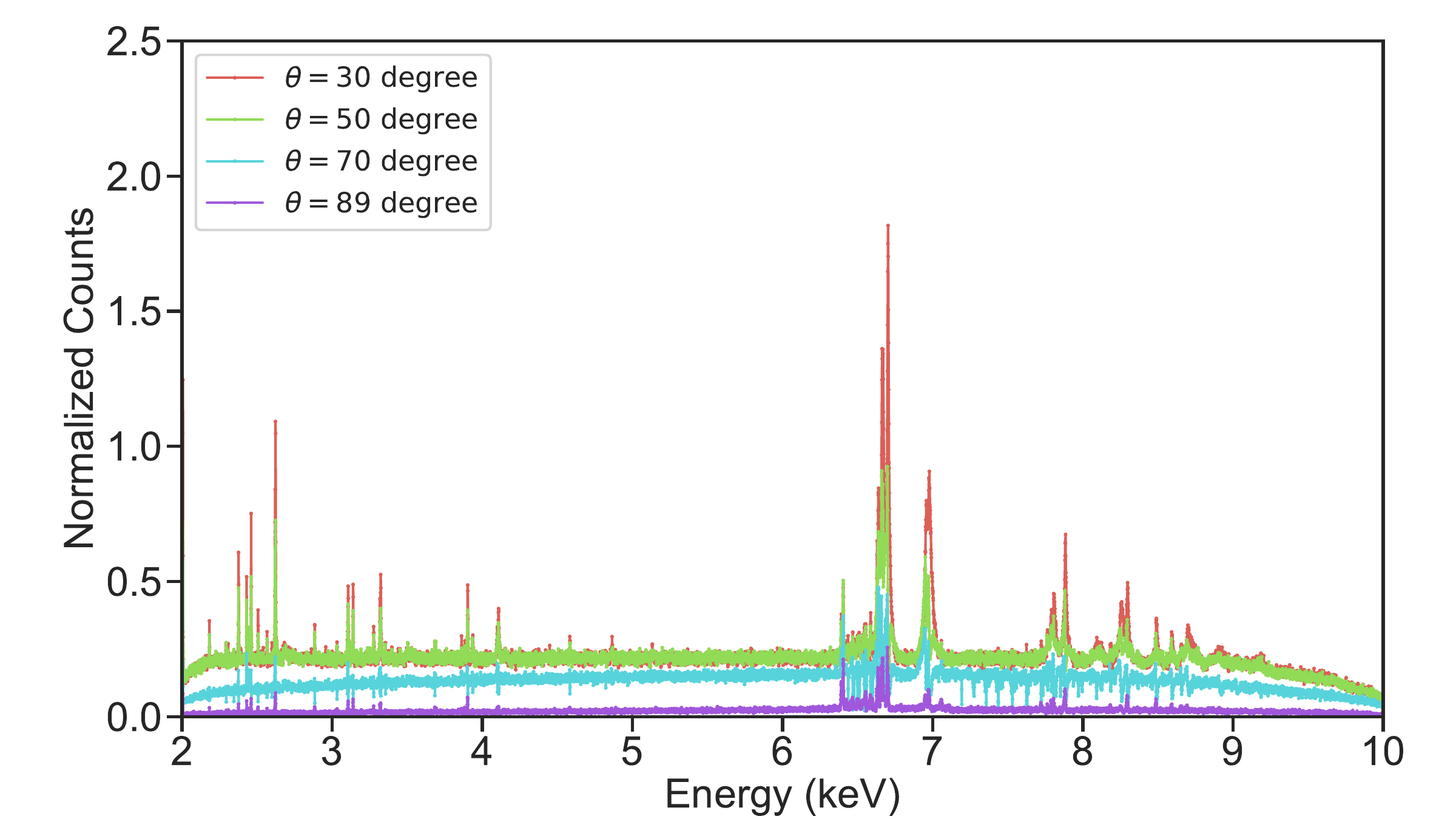}{0.5\textwidth}{(a)}\fig{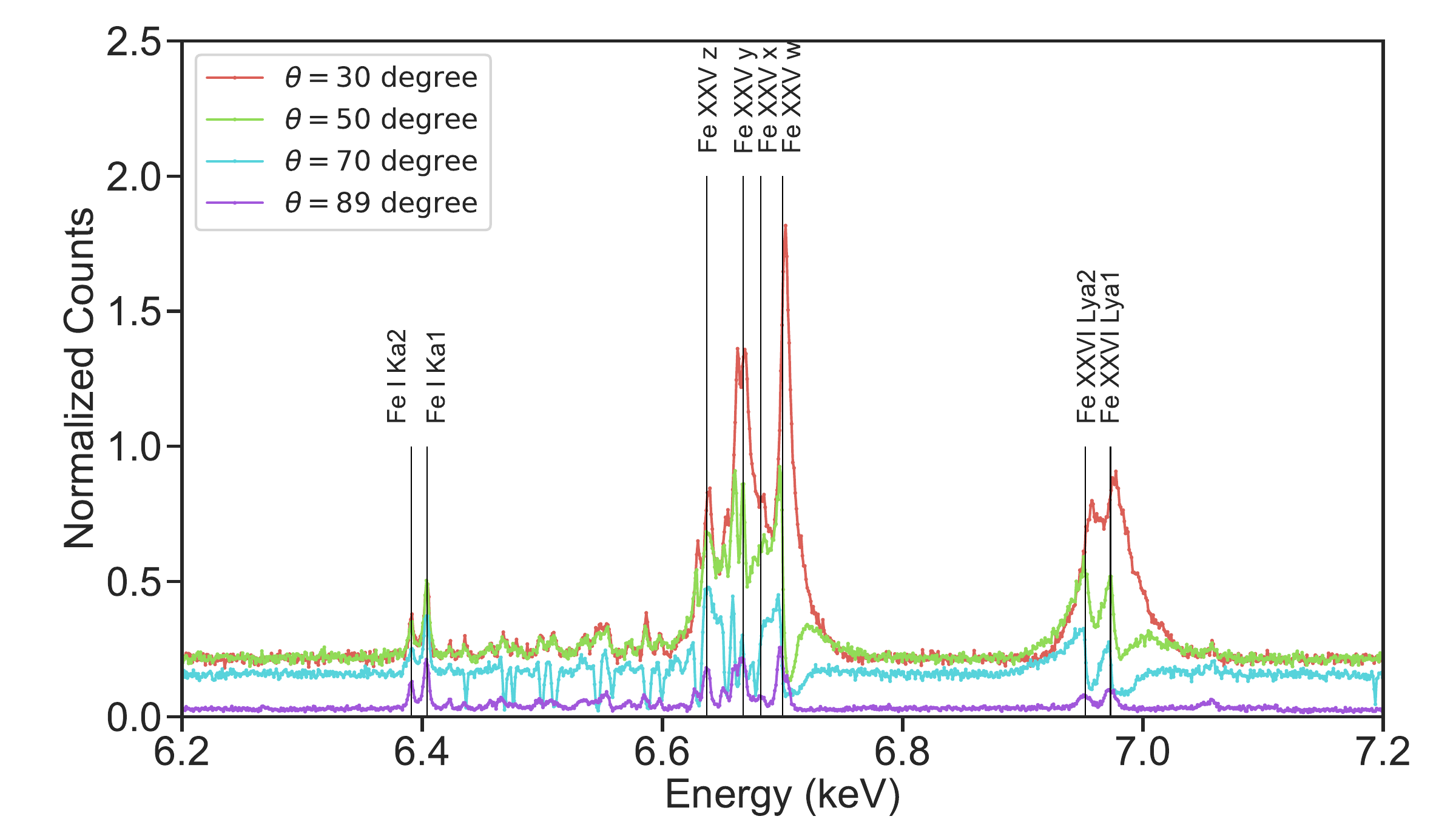}{0.5\textwidth}{(b)}}
\caption{(a) The scattered component of the X-ray spectrum at $2$--$10 \ \mathrm{keV}$ in the rest frame. The red, green, light blue, and purple line show the X-ray spectrum at the polar angle $\theta = 30\degr$, $50\degr$, $70\degr$, and $89\degr$, respectively. (b) The scattered component of the X-ray spectrum at $6.2$--$7.2 \ \mathrm{keV}$. The black lines correspond to the energy of the emission lines in the rest frame due to \ion{Fe}{1} K$\alpha$2 ($6.391 \ \mathrm{keV}$), \ion{Fe}{1} K$\alpha$1 ($6.404 \ \mathrm{keV}$), \ion{Fe}{25} z ($6.637 \ \mathrm{keV}$), \ion{Fe}{25} y ($6.667 \ \mathrm{keV}$), \ion{Fe}{25} x ($6.682 \ \mathrm{keV}$), \ion{Fe}{25} w ($6.700 \ \mathrm{keV}$), \ion{Fe}{26} Ly$\alpha$2 ($6.952 \ \mathrm{keV}$), and \ion{Fe}{26} Ly$\alpha$1 ($6.973 \ \mathrm{keV}$). Energies of absorption lines are based on the Flexible Atomic Code \citep{Gu2008}.}
\end{figure*}
\begin{figure*}\label{Figure0008}
\gridline{\fig{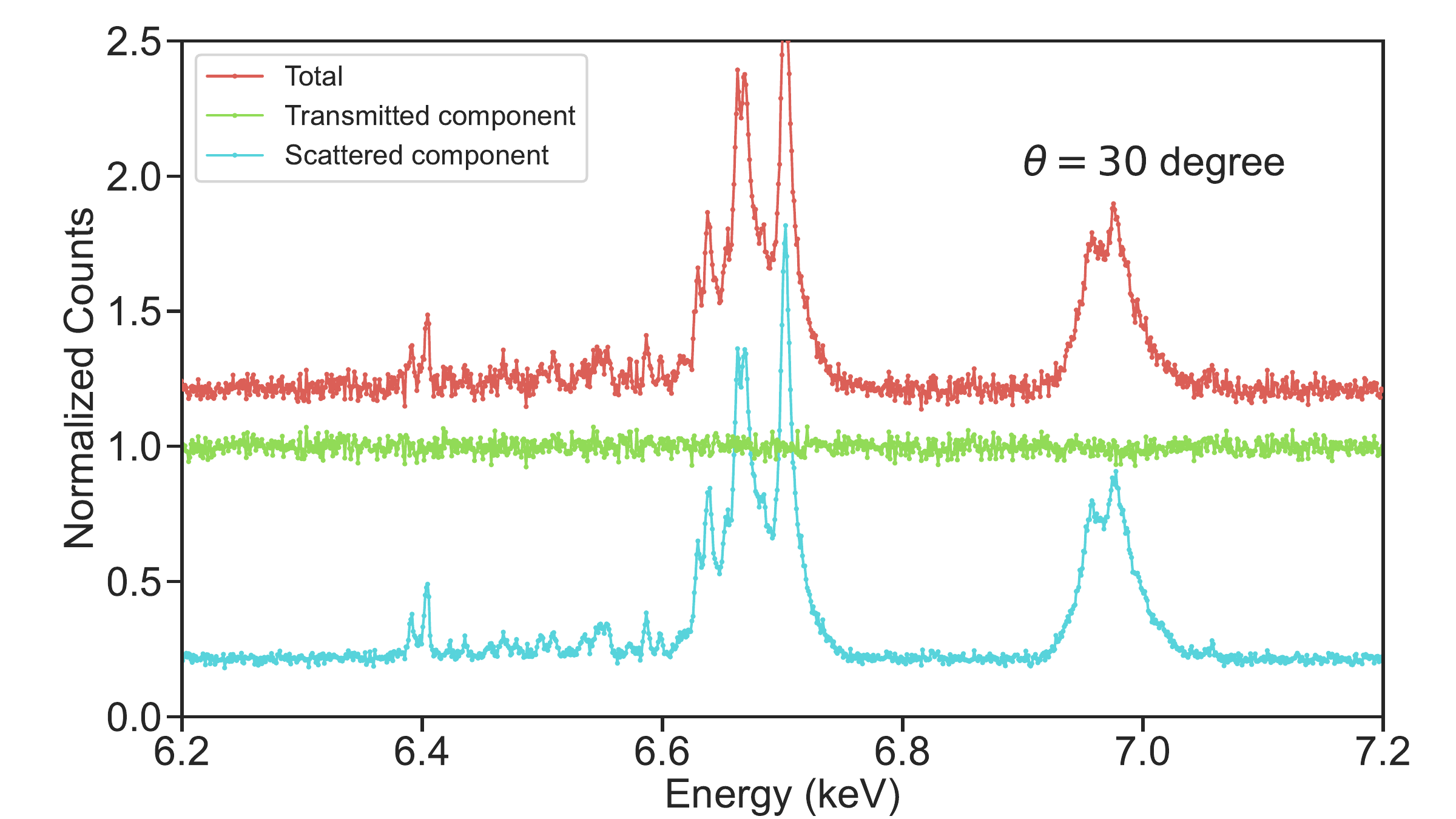}{0.5\textwidth}{(a)}\fig{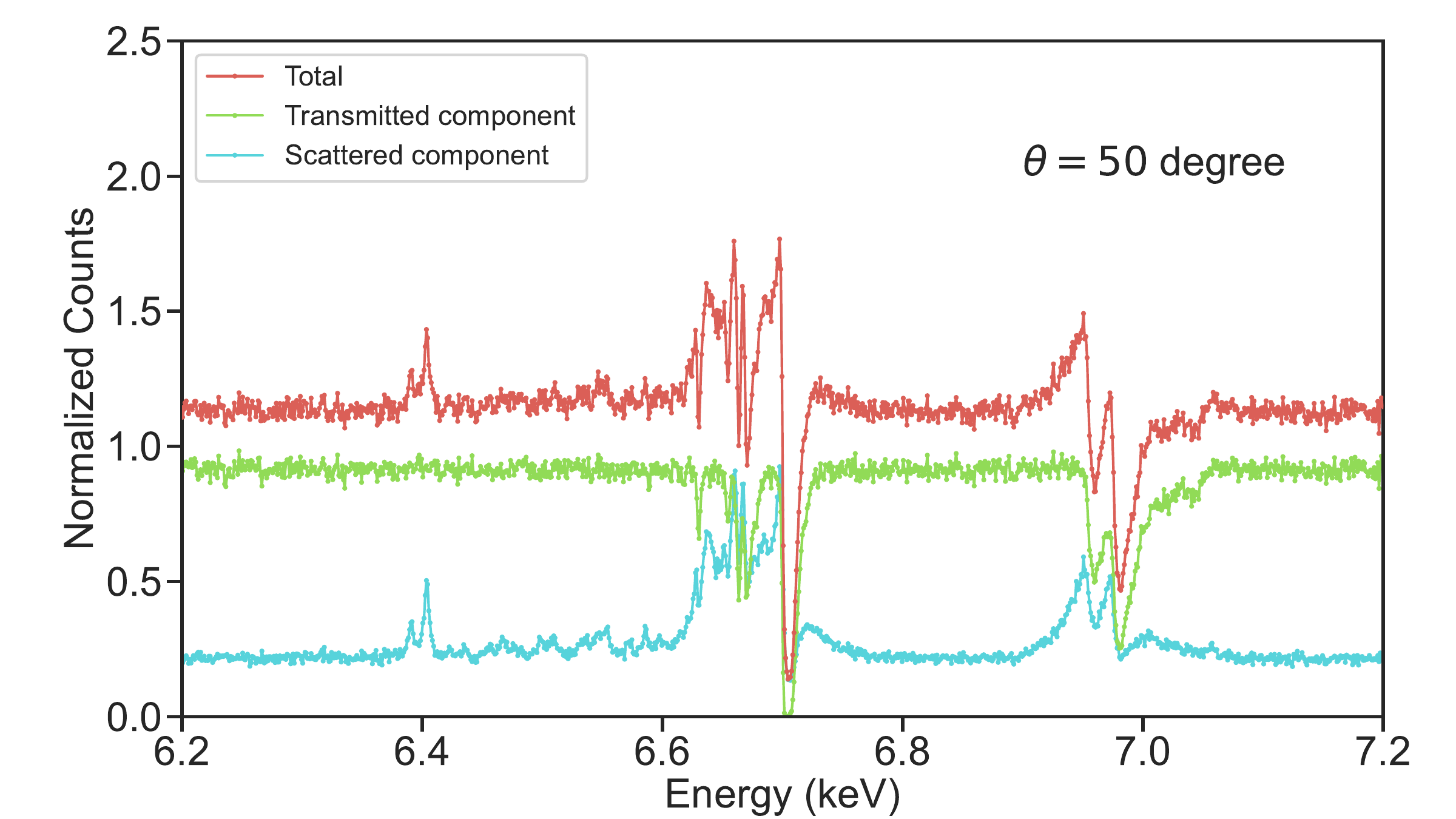}{0.5\textwidth}{(b)}}
\gridline{\fig{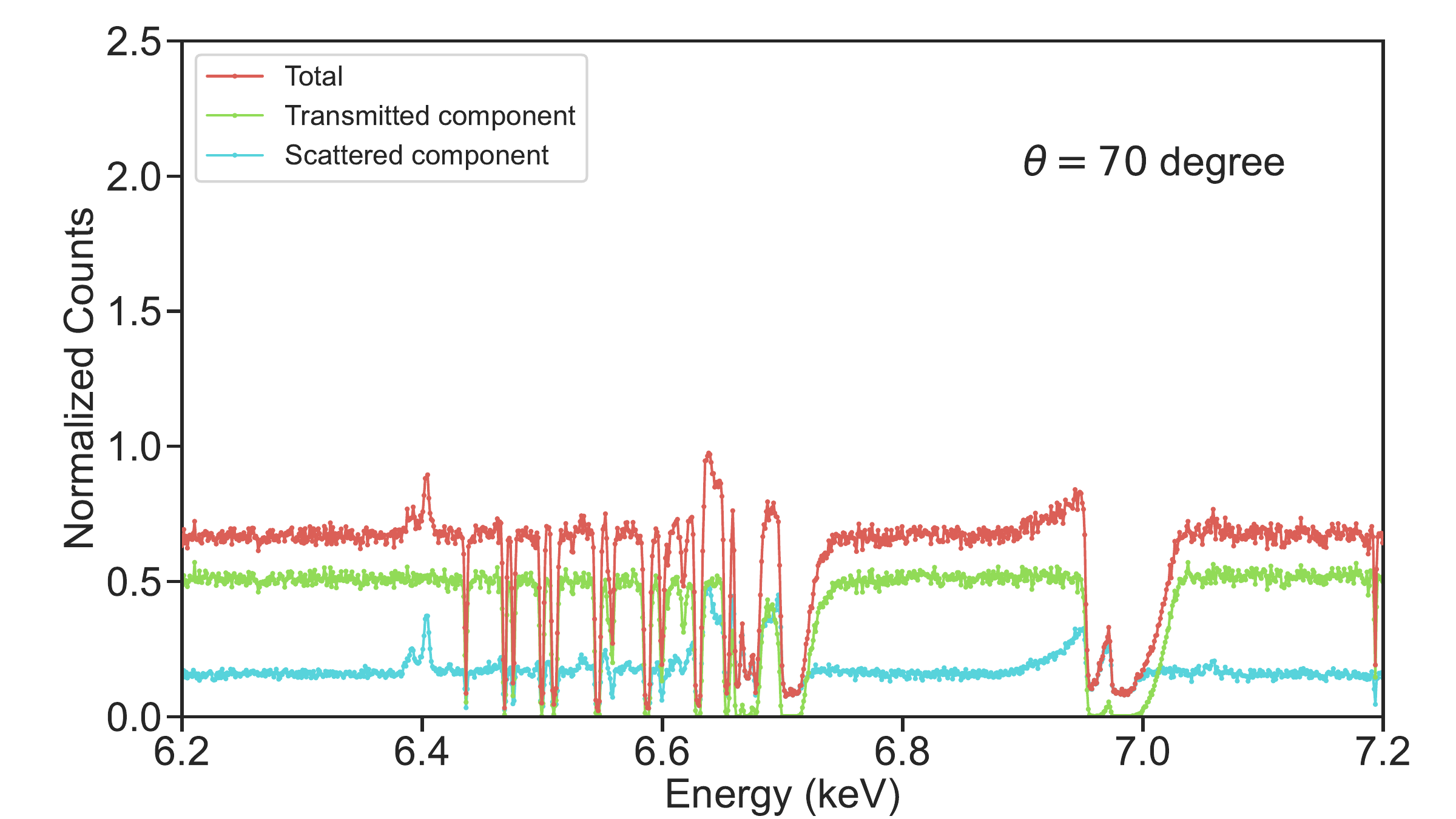}{0.5\textwidth}{(c)}\fig{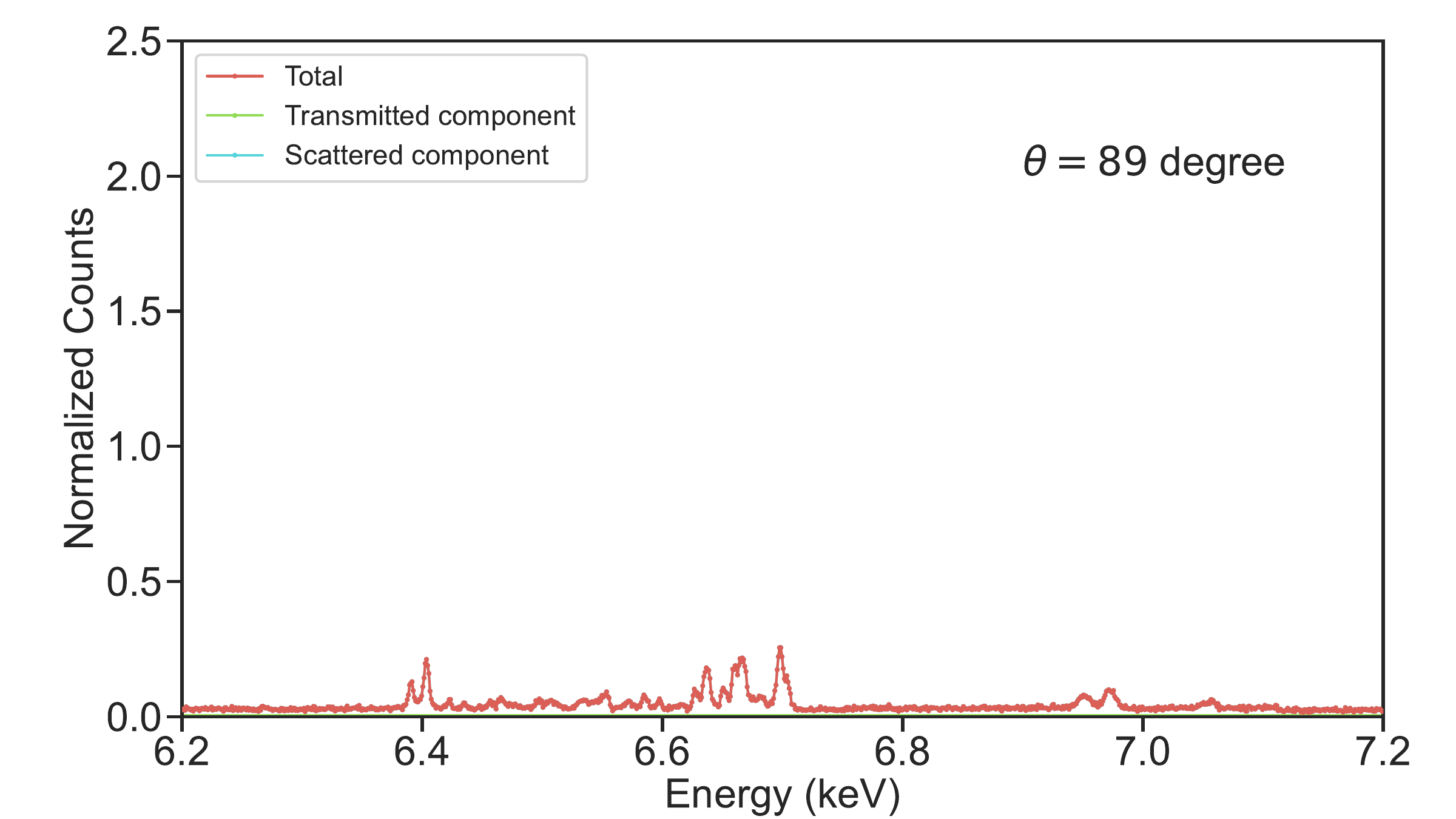}{0.5\textwidth}{(d)}}
\caption{(a) The total X-ray spectra at the polar angle $\theta = 30\degr$. (b) The total X-ray spectra at the polar angle $\theta = 50\degr$. (c) The total X-ray spectra at the polar angle $\theta = 70\degr$. (d) The total X-ray spectra at the polar angle $\theta = 89\degr$. P-Cygni-like feature is seen in total spectrum.}
\end{figure*}\clearpage
\section{Results}\label{Section0300}
We present the results obtained from the calculations. \hyperref[Section0301]{Section~3.1} shows the distributions of the hydrogen number density and the radial velocity based on the wind model. \hyperref[Section0302]{Section~3.2} presents the distribution of the turbulent velocity, the temperature, and the ionization parameter. \hyperref[Section0303]{Section~3.3} shows the transmitted component and the scattered component obtained from the X-ray radiative transfer calculation.
\subsection{Self-similar Solution of Magnetically Driven Disk Wind}\label{Section0301}
We calculated the hydrogen number density and the three-dimensional velocities of the disk wind under steady-state and axisymmetric assumptions. \hyperref[Figure0004]{Figure~4(a)} shows the distribution of the hydrogen number density. The hydrogen number density decreases with radius, while it increases with the polar angle.  \hyperref[Figure0004]{Figure~4(b)} shows the distribution of the radial velocity. Radial velocity decreases with radius and polar angle, as expected, according to \hyperref[Equation0001]{Equation~1}. In other words, the blueshift of the observed absorption lines is expected to be larger at smaller distances. \added{Unlike thermal-radiative winds, we emphasize that magnetically driven winds can be launched from smaller disk radii even within Compton radius \citep[e.g.][]{Tomaru2020}, covering a large radial extent of the disk}.
\subsection{Photoionized Equilibrium Calculations}\label{Section0302}
We calculated the photoionized equilibrium based on the density distributions. \hyperref[Figure0004]{Figure~4(c)} shows the temperature distribution. The temperature decreases with the radius and polar angle. \hyperref[Figure0004]{Figure~4(d)} shows the turbulent velocity distribution. The turbulent velocity is mainly determined by the shear velocity in regions where the wind speed is near-relativistic, while by the speed of sound where the wind speed is slower and the gas is still sufficiently hot. \hyperref[Figure0005]{Figure~5} shows the distribution of the ionization parameter. The ionization parameter decreases with the radius and polar angle. Since the turbulent velocity also decreases with radius and polar angle, the widths of the observed absorption lines are expected to be narrower for absorption lines of less ionized ions.
\subsection{X-Ray Radiative Transfer}\label{Section0303}
\subsubsection{Transmitted Component of X-Ray Spectrum}
As a result of X-ray radiative transfer calculations, we obtain various spectral features from a series of ions. \hyperref[Figure0006]{Figure~6(a)} shows the dependence of the transmitted component of the X-ray spectrum at $2$--$10 \ \mathrm{keV}$ in the rest frame on the polar angle. If the polar angle is $30\degr$, the transmitted component does not show absorption lines. This is because the material along the line of sight is completely ionized. When the polar angle is between $50\degr$ and $70\degr$, the transmitted component exhibits absorption lines due to the photoionized material whose ionization parameter is $10^2$--$10^4 \ \mathrm{erg} \ \mathrm{cm} \ \mathrm{s}^{-1}$. In particular, the polar angle of $70\degr$, since the ionization parameter decreases with the polar angle, the absorption lines are observed due to less ionized materials. If the polar angle is $89\degr$, we cannot observe the transmitted component due to the presence of a Compton-thick material in the line of sight.

To focus on the Fe K complex, we zoomed in on the $6.2$--$7.2 \ \mathrm{keV}$ range. \hyperref[Figure0006]{Figure~6(b)} shows the transmitted component of the X-ray spectrum at $6.2$--$7.2 \ \mathrm{keV}$. The black lines correspond to the energy of the absorption lines in the rest frame due to \added{\ion{Fe}{19} ($6.468 \ \mathrm{keV}$), \ion{Fe}{20} ($6.508 \ \mathrm{keV}$), \ion{Fe}{21} ($6.545 \ \mathrm{keV}$), \ion{Fe}{22} ($6.586 \ \mathrm{keV}$), \ion{Fe}{23} ($6.628 \ \mathrm{keV}$), \ion{Fe}{24} ($6.661 \ \mathrm{keV}$), \ion{Fe}{25} ($6.700 \ \mathrm{keV}$), \ion{Fe}{26} Ly$\alpha$2 ($6.952 \ \mathrm{keV}$), and \ion{Fe}{26} Ly$\alpha$1 ($6.973 \ \mathrm{keV}$)}. If the polar angle is $50\degr$, we detected the blueshifted absorption lines of \ion{Fe}{24} ($6.663 \ \mathrm{keV}$), \ion{Fe}{25} ($6.703 \ \mathrm{keV}$), \ion{Fe}{26} Ly$\alpha$2 ($6.959 \ \mathrm{keV}$), and \ion{Fe}{26} Ly$\alpha$1 ($6.980 \ \mathrm{keV}$). These correspond to the speed of $90 \ \mathrm{km} \ \mathrm{s}^{-1}$, $130 \ \mathrm{km} \ \mathrm{s}^{-1}$, and $300 \ \mathrm{km} \ \mathrm{s}^{-1}$, respectively. This result indicates that these absorption lines are generated in different parts of the wind. We study the number density distributions of \ion{Fe}{23}, \ion{Fe}{24}, \ion{Fe}{25}, and \ion{Fe}{26} and discuss the origin of these absorption lines in \hyperref[Section0401]{Section~4.1}. \hyperref[Figure0006]{Figure~6(b)} also demonstrates that the absorption line profile differs for different charge states. For example, the profile of the \ion{Fe}{25} absorption line is an almost symmetric Gaussian-like shape, while that of \ion{Fe}{26} is asymmetric and has a tail on the blue side. This result is consistent with some previous work \citep[e.g.,][]{Fukumura2022, Gandhi2022, Chakravorty2023, Datta2024} in which it is suggested that line profiles can be noticeably distinct among various wind driving mechanisms. The profile of the \ion{Fe}{26} absorption line shows a red tail in the case of UV-line-driven disk winds, whereas it shows a blue tail in the case of magnetically driven disk winds.

\clearpage
\subsubsection{Scattered Component of X-Ray Spectrum}
Since MONACO is a three-dimensional Monte Carlo X-ray radiative transfer code, it can accurately handle the scattered component. \hyperref[Figure0007]{Figure~7(a)} shows the dependence of the scattered component of the X-ray spectrum at $2$--$10 \ \mathrm{keV}$ in the rest frame on the polar angle. Although the scattered component also depends on the polar angle, its dependence is less than that of the transmitted component. This is because the scattered component originates from various regions (see also \citealt{Fukumura2025}).

Having been focused again on the Fe K band, \hyperref[Figure0007]{Figure~7(b)} shows the scattered component of the X-ray spectrum at $6.2$--$7.2 \ \mathrm{keV}$. The black lines correspond to the energy of the emission lines in the rest frame of \added{\ion{Fe}{1} K$\alpha$2 ($6.391 \ \mathrm{keV}$), \ion{Fe}{1} K$\alpha$1 ($6.404 \ \mathrm{keV}$), \ion{Fe}{25} z ($6.637 \ \mathrm{keV}$), \ion{Fe}{25} y ($6.667 \ \mathrm{keV}$), \ion{Fe}{25} x ($6.682 \ \mathrm{keV}$), \ion{Fe}{25} w ($6.700 \ \mathrm{keV}$), \ion{Fe}{26} Ly$\alpha$2 ($6.952 \ \mathrm{keV}$), and \ion{Fe}{26} Ly$\alpha$1 ($6.973 \ \mathrm{keV}$)}. If the polar angle is $30\degr$, the scattered component shows emission lines such as \ion{Fe}{1} K$\alpha$, \ion{Fe}{25} z, y, x, w, and \ion{Fe}{26} Ly$\alpha$. The peak energy of \ion{Fe}{26} Ly$\alpha$2 is $6.957 \ \mathrm{keV}$ and that of \ion{Fe}{26} Ly$\alpha$1 is $6.977 \ \mathrm{keV}$. This suggests that the emission lines of are blueshifted with a velocity of approximately $200 \ \mathrm{km} \ \mathrm{s}^{-1}$. When the polar angle is $50\degr$, the scattered component exhibits the absorption lines of \ion{Fe}{25} and \ion{Fe}{26}, in addition to these emission lines.

\subsubsection{Total X-Ray Spectrum}
What we can actually observe is the sum of the transmitted component and the scattered component. \hyperref[Figure0008]{Figure~8} shows the total X-ray spectrum at (a) the polar angle $\theta = 30\degr$, (b) $\theta = 50\degr$, (c) $\theta = 70\degr$, and (d) $\theta = 89\degr$. It is clear that the overall nature of the expected spectrum is primarily characterized by the polar angle; i.e. varying from absorption-dominated spectrum for high polar angle to emission-dominated spectrum for low polar angle. To be specific, \hyperref[Figure0008]{Figure~8(a)} shows that the total spectrum exhibits only emission lines. In \hyperref[Figure0008]{Figure~8(b)}, absorption lines due to \ion{Fe}{25} and \ion{Fe}{26} appear in addition to the remaining emission lines. Note that a superposition of such components naturally imprints a P-Cygni profile. \hyperref[Figure0008]{Figure~8(c)} shows that the total spectrum consists of a series of rich absorption signatures over a wide range of ionization states such as \ion{Fe}{19}, \ion{Fe}{20}, \ion{Fe}{21}, \ion{Fe}{22}, \ion{Fe}{23}, \ion{Fe}{24}, \ion{Fe}{25}, and \ion{Fe}{26}. The same absorption lines seen in \hyperref[Figure0008]{Figure~8(b)} are still present, but their strength is determined by the scattered component. This may affect the equivalent width (EW) of the predicted absorption lines. On the other hand, in \hyperref[Figure0008]{Figure~8(d)}, the plasma is generally optically thick near the base of the disk wind, allowing a substantial scattering event to occur. As a result, scattered photons can partially fill in the absorption features. Thus, we can practically expect to observe emission-dominated spectrum with strong scattered component. These characteristic spectral features, which are sensitive to wind density due to polar angle, can uniquely provide a useful diagnostic proxy to unveil the underlying wind condition.\clearpage
\begin{figure*}\label{Figure0009}
\gridline{\fig{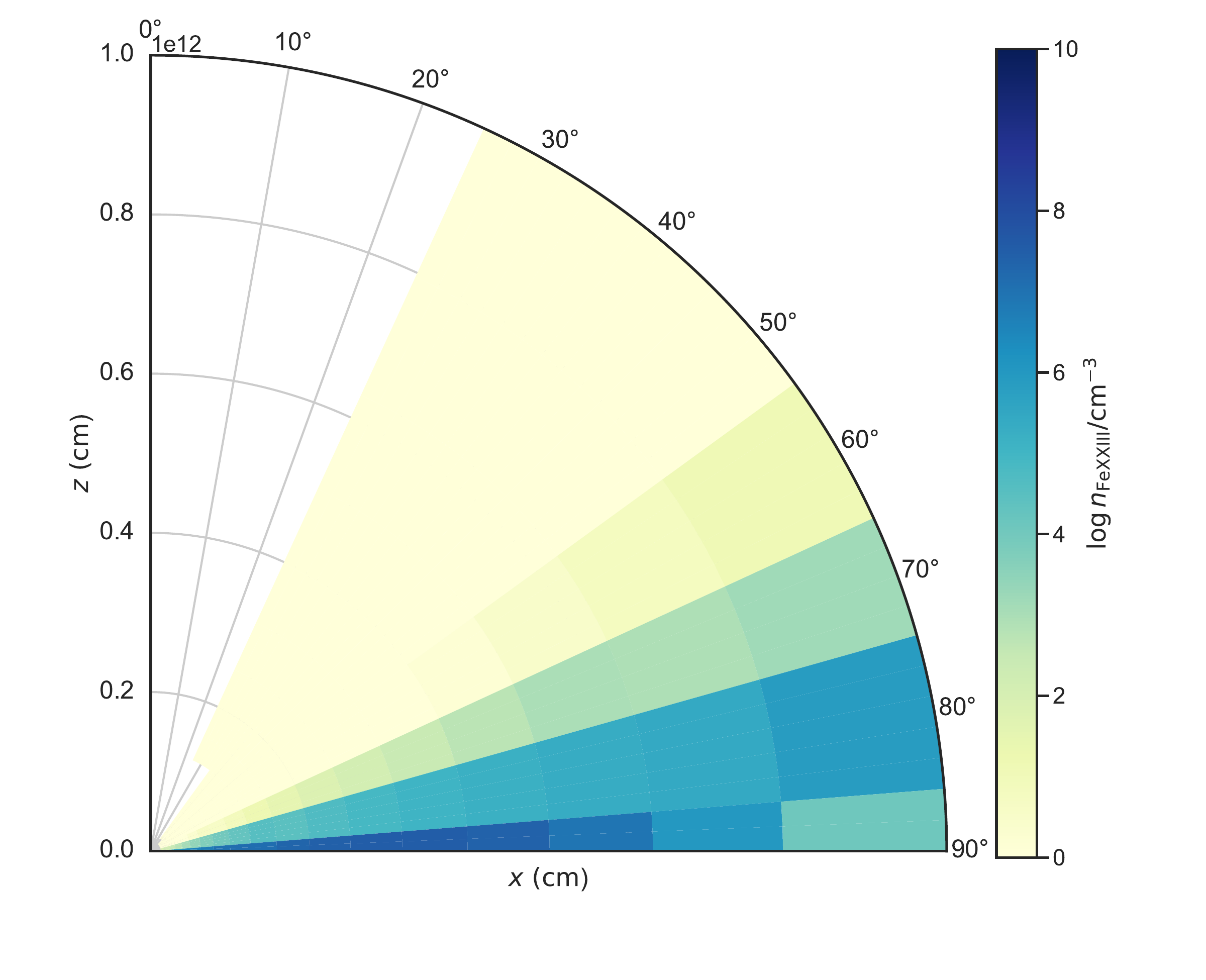}{0.5\textwidth}{(a)}\fig{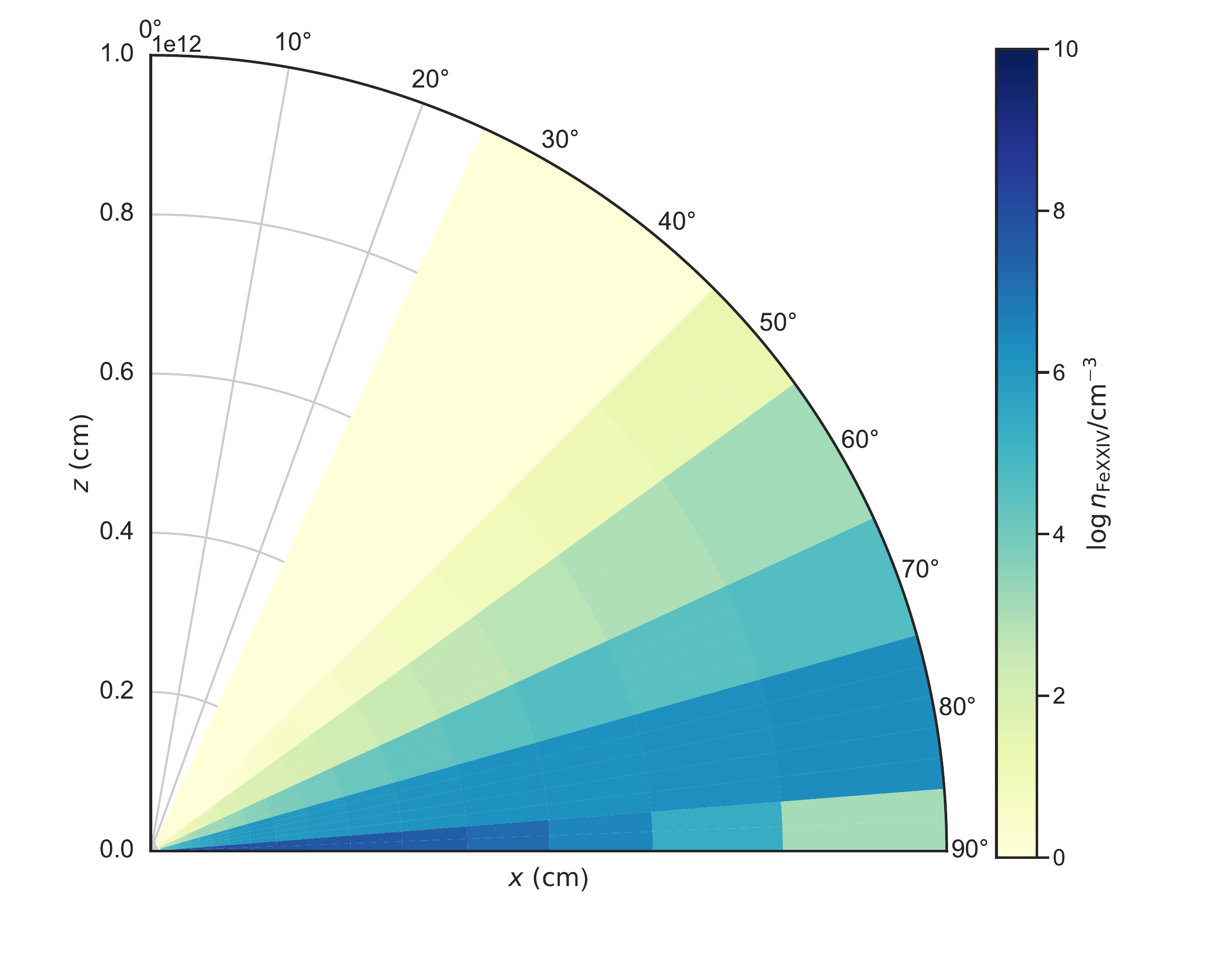}{0.5\textwidth}{(b)}}
\gridline{\fig{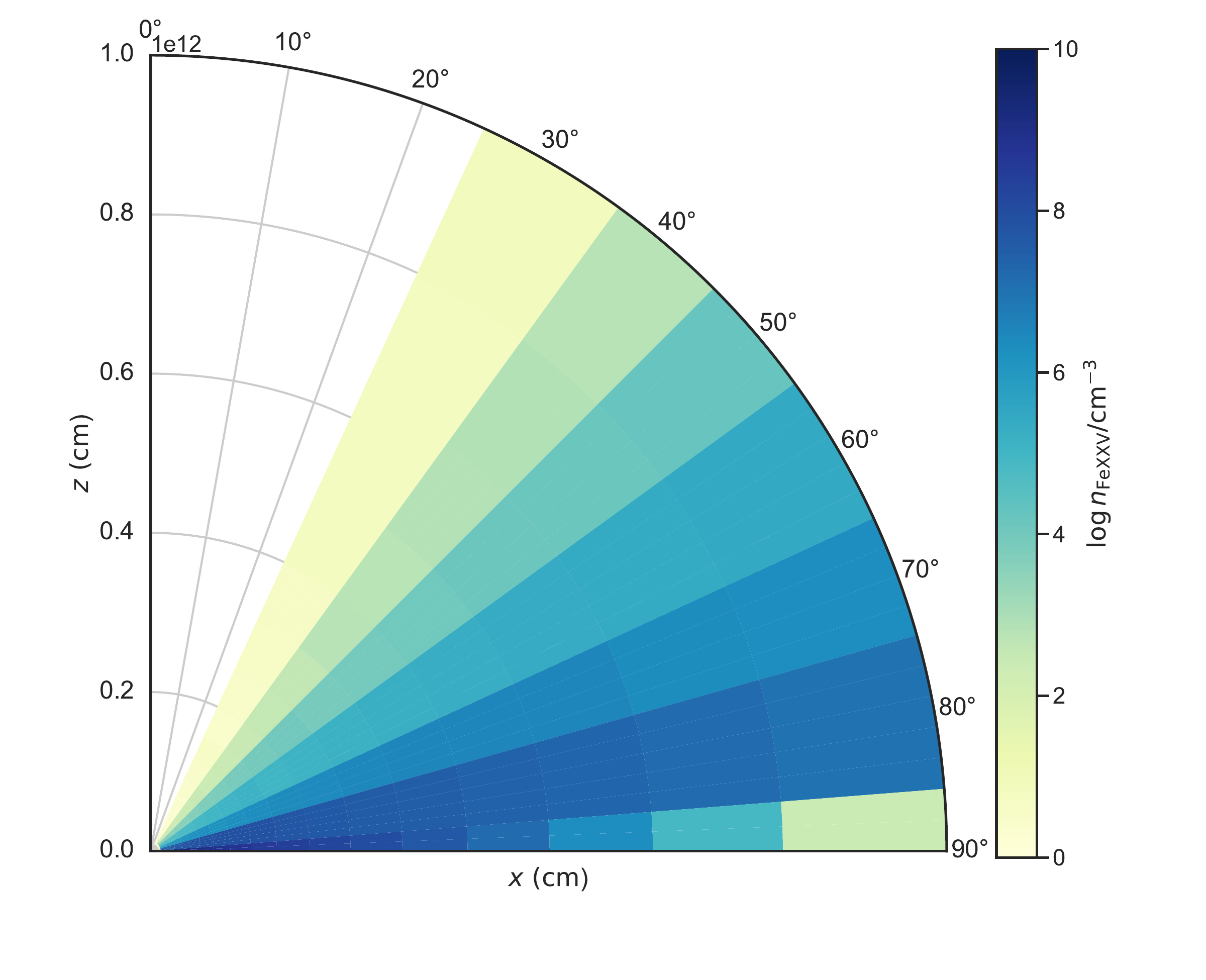}{0.5\textwidth}{(c)}\fig{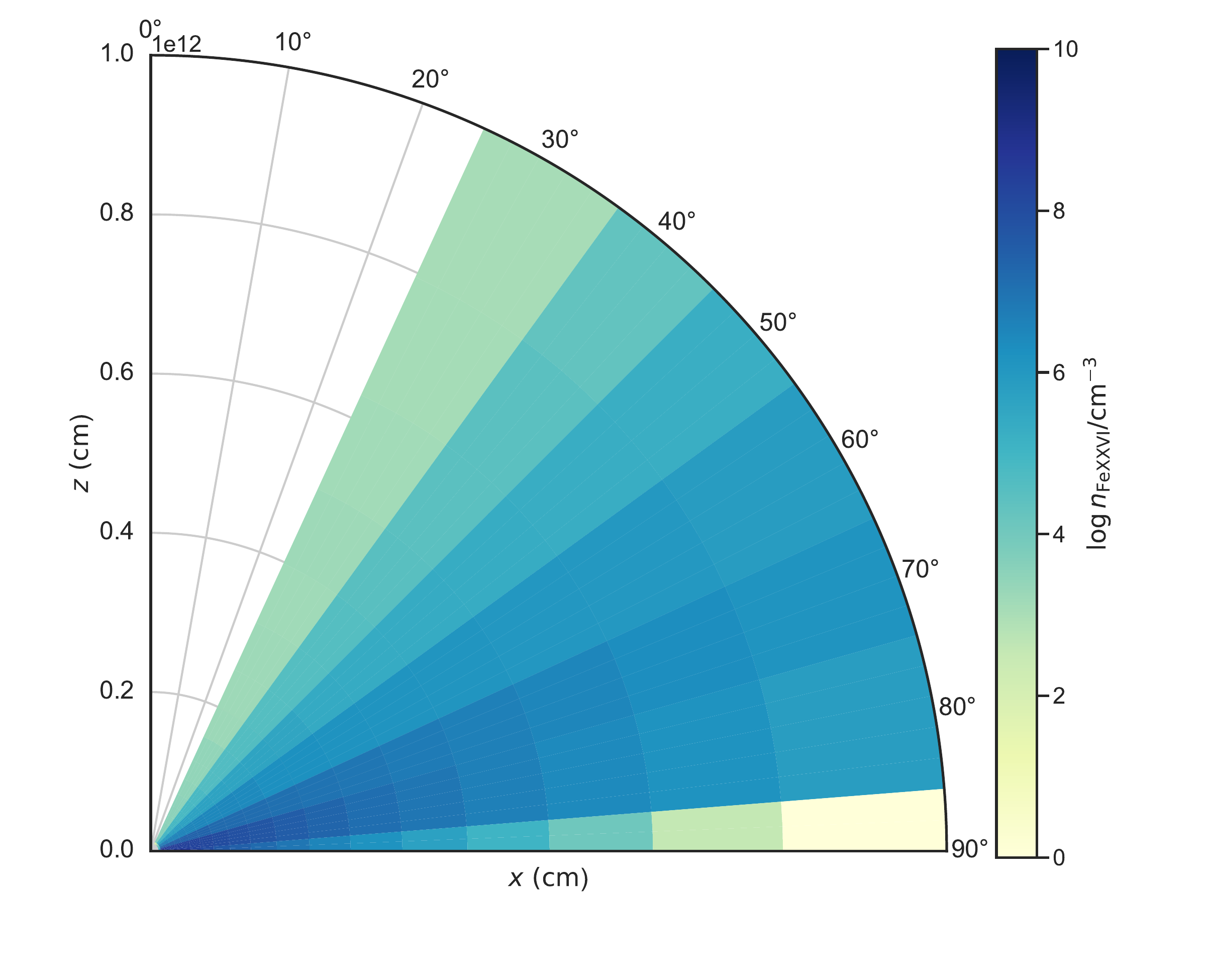}{0.5\textwidth}{(d)}}
\caption{(a)~The distribution of the \ion{Fe}{23} number density $\log n_{\mathrm{Fe XXIII}}/\mathrm{cm}^{-3}$. (b)~The distribution of the \ion{Fe}{24} number density $\log n_{\mathrm{Fe XXIV}}/\mathrm{cm}^{-3}$. (c)~The distribution of the \ion{Fe}{25} number density $\log n_{\mathrm{Fe XXV}}/\mathrm{cm}^{-3}$. (d)~The distribution of the \ion{Fe}{26} number density $\log n_{\mathrm{Fe XXVI}}/\mathrm{cm}^{-3}$.}
\end{figure*}
\begin{figure*}\label{Figure0010}
\gridline{\fig{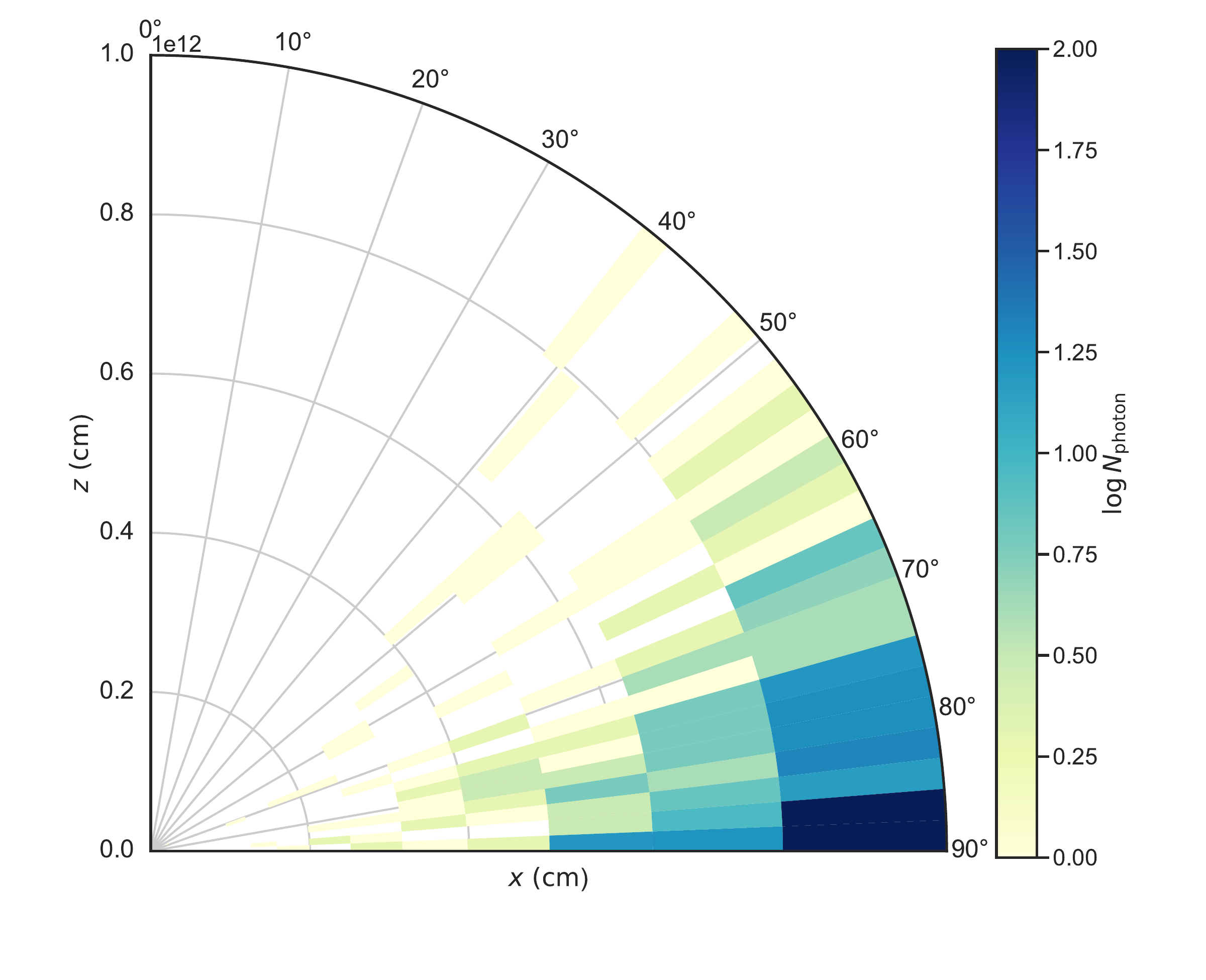}{0.5\textwidth}{}}
\caption{The distribution of the number of neutral iron fluorescent emission lines emitted.}
\end{figure*}
\begin{figure*}\label{Figure0011}
\gridline{\fig{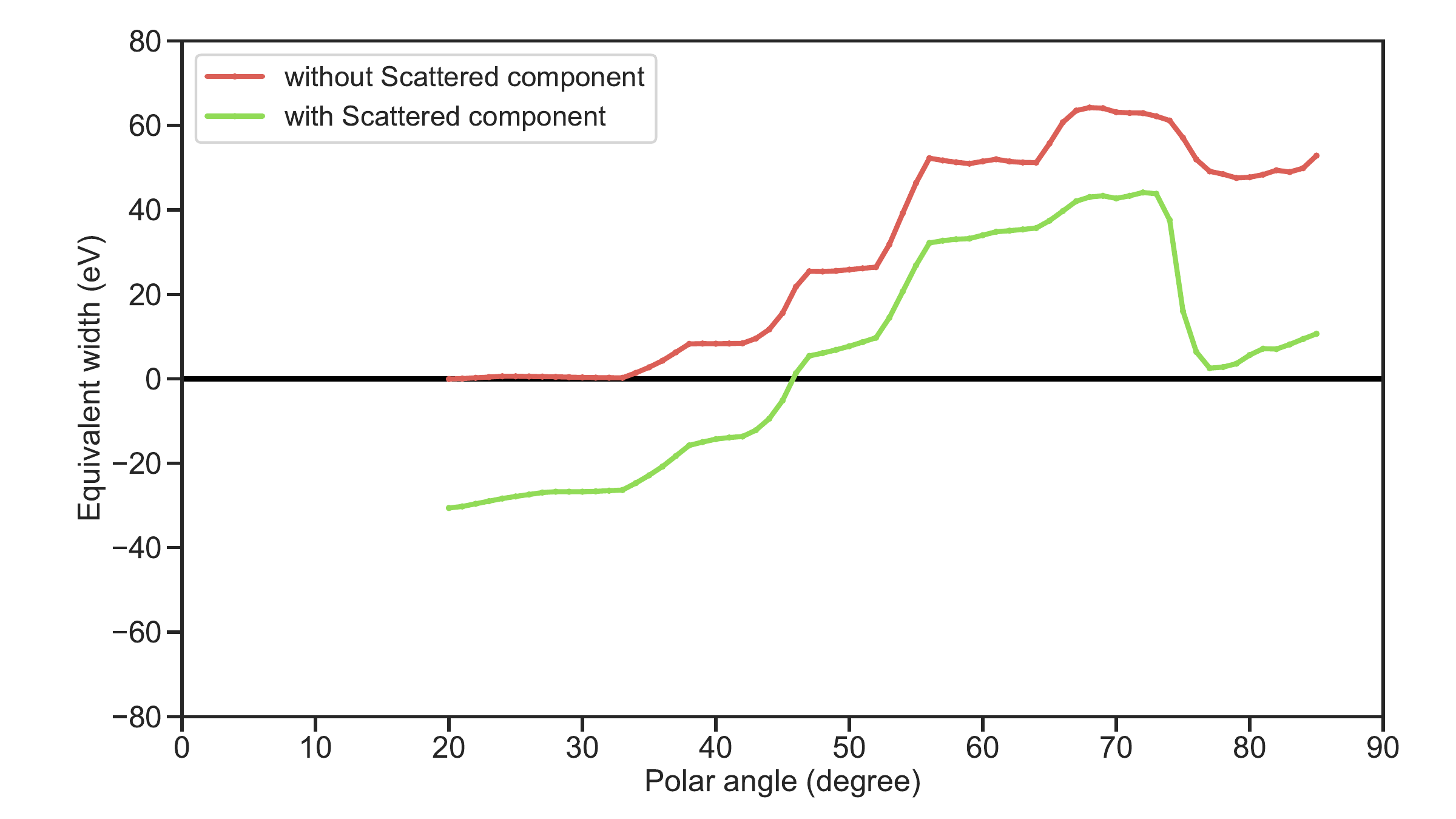}{0.5\textwidth}{}}
\caption{The equivalent width (EW) of absorption and emission lines of \ion{Fe}{26} as a function of the polar angle. The black lines represent $\mathrm{EW} = 0$. $\mathrm{EW} < 0$ represents the emission line, while $\mathrm{EW} > 0$ corresponds to the absorption line.}
\end{figure*}\clearpage
\section{Discussion}\label{Section0400}
\added{In this study, we calculated transmitted and scattered components of X-ray spectrum and investigated the effect of scattering in the wind on the spectral features as a function of the polar angle. We find that the amount of contributions of these individual components to the predicted spectra is very sensitive to the polar angle, as expected; that is, an absorption-dominated spectrum for a high polar angle and an emission-dominated spectrum for a low polar angle. A well-defined P-Cygni structure is also found in our simulations for an intermediate polar angle. In particular, the calculated absorption line profiles are uniquely skewed with an extended blue tail of asymmetry, as a characteristic feature of the magnetically driven disk wind, as predicted before. \hyperref[Section0401]{Section~4.1} discusses the origin of the absorption lines. \hyperref[Section0402]{Section~4.2} presents the origin of the neutral iron fluorescent lines. \hyperref[Section0403]{Section~4.3} discusses the effect of scattering. \hyperref[Section0404]{Section~4.4} presents the P-Cygni features}.
\subsection{The Origin of Absorption Lines}\label{Section0401}
\added{To determine where absorption lines are produced, we examined the distribution of the number density of \ion{Fe}{23}, \ion{Fe}{24}, \ion{Fe}{25}, and \ion{Fe}{26}. \hyperref[Figure0009]{Figure~9} shows the distributions of (a) \ion{Fe}{23} number density, (b) \ion{Fe}{24} number density, (c) \ion{Fe}{25} number density and (d) \ion{Fe}{26} number density. \hyperref[Figure0009]{Figure~9(a)--(b)} indicate that \ion{Fe}{23} and \ion{Fe}{24} are abundant in the region of the polar angle of $60\degr$ to $80\degr$. On the other hand, \hyperref[Figure0009]{Figure~9(c)--(d)} shows that \ion{Fe}{25} and \ion{Fe}{26} are abundant in the region of the polar angle of $50\degr$ to $80\degr$. We note that although \ion{Fe}{23} and \ion{Fe}{24} exist in the region where the polar angle is greater than $85\degr$, these absorption lines are hard to observe due to the presence of Compton-thick materials outside of this region}. 

The profile of the \ion{Fe}{26} absorption lines is uniquely characterized by an extended blue tail of asymmetry. The driving mechanism of the disk wind can be distinguished by measuring the profile of the \ion{Fe}{26} absorption lines. The X-Ray Imaging and Spectroscopy Mission (XRISM; \citealt{Tashiro2025}) is equipped with a microcalorimeter detector (Resolve; \citealt{Ishisaki2022}). Since the energy resolution of Resolve at $6 \ \mathrm{keV}$ is approximately $30$ times higher than that of the CCD detectors, Resolve can accurately capture the profile of \ion{Fe}{26} absorption lines including the doublet structure at microcalorimeter resolution, which would otherwise be very challenging even with dispersive instruments \citep[e.g.,][]{Miller2015a, Tsujimoto2025}. Note that a different velocity field of winds would be reflected in a distinct line profile shape. Our primary focus in the present work is to examine the effects of the scattering process for various polar angles in the context of MHD-driven disk winds.
\subsection{The Origin of the Neutral Iron Fluorescent Emission Lines}\label{Section0402}
Since MONACO keeps track of where and what events occur, we look for where the \ion{Fe}{1} fluorescence emission lines are emitted. \hyperref[Figure0010]{Figure~10} shows the distribution of the number of \ion{Fe}{1} fluorescence emission lines emitted. \hyperref[Figure0010]{Figure~10} indicates that \ion{Fe}{1} fluorescence emission lines are emitted mainly from regions with an polar angle greater than $70\degr$. Since the Keplerian rotational velocity at a radius of $10^{12} \ \mathrm{cm}$ is $300 \ \mathrm{km} \ \mathrm{s}^{-1}$, the \ion{Fe}{1} fluorescent emission line has a width of approximately $6 \ \mathrm{eV}$. In this case, we can observe the \ion{Fe}{1} K$\alpha2$ and K$\alpha1$ separately, as shown in \hyperref[Figure0006]{Figure~6(b)}.\clearpage
\subsection{Effect of Scattering}\label{Section0403}
The strength of the atomic features can be quantitatively estimated by the equivalent width (EW) defined as
\begin{equation}
EW \equiv \int_{E_{\mathrm{min}}}^{E_\mathrm{max}} \frac{F_{\mathrm{cont}} - F_{\mathrm{total}}}{F_{\mathrm{cont}}} dE
\end{equation}
where, $E_{\mathrm{min}}$ is the minimum energy, $E_{\mathrm{max}}$ is the maximum energy, $F_{\mathrm{cont}}$ is the flux of the continuum and $F_{\mathrm{total}}$ is the total flux over an absorption energy band. The positive EW represent the absorption lines, and the EW widths represent the emission lines. As shown in \added{\hyperref[Figure0008]{Figure~8}}, the depths of the absorption lines are affected by the scattering component.

We compared the EW of the transmitted component with that of the total. \hyperref[Figure0011]{Figure~11} shows the EW of the \ion{Fe}{26} resonance transition as a function of the polar angle. The EW of the transmitted component increases with the polar angle below $70\degr$, while it decreases with the polar angle above $70\degr$ and is almost constant. This is because the \ion{Fe}{26} absorption lines are generally saturated if the polar angle is greater than $70\degr$. The EW with the scattered component shows almost the same trend as that of the transmitted component. However, the EW with the scattering component is systematically smaller than that without scattering by $20$--$50 \ \mathrm{eV}$. Since the ratio of the scattered component to the total is higher, these differences are especially large when the polar angle is greater than $70\degr$. Thus, it is very important to consider the effect of scattering to better estimate the intrinsic wind property.

To follow up on theoretical implications associated with scattering processes considered in this work, a couple of plausible situations can be considered in the context of BH XRBs. First, it has been seen that the BH transients can be episodically obscured by multiple zones of ionized gas ($\log \xi/\mathrm{erg} \ \mathrm{cm} \ \mathrm{s}^{-1} \sim 3-4$) with high column densities ($N_{\mathrm{H}} \sim 10^{23} \ \mathrm{cm}^{-2}$) possibly produced by a failed wind \citep[e.g.,][for the recent dim state of GRS~1915+105]{Miller2020}. This obscured state of BH XRBs may be naturally explained in our model with a high density $n_{H}$ in \hyperref[Equation0001]{Equation~1} with a high polar angle as in GRS~1915+105. Our calculations demonstrate that the scattered component plays a dominant role in characterizing the total spectrum as absorption features in the transmitted component are effectively filled in by scattered photons, especially at the base of the wind (\hyperref[Figure0008]{Figure~8}).

Second, it is suggested that there exists another class of disk winds in BH XRBs; highly ionized fast outflows (i.e., $v \sim 0.07c$) of massive column (i.e., $N_{\mathrm{H}} \simeq 10^{23} \ \mathrm{cm}^{-2}$), in low-polar angle ($\theta \lsim 30\degr-60\degr$) BH XRBs \citep{Chakraborty2021, Chakraborty2025}. If confirmed, these massive winds could effectively participate in the scattering process near the polar region, perhaps making an appreciable contribution to the equatorial, high-polar angle portion of the wind. For example, intrinsically deep (saturated) absorption features (e.g., \ion{Fe}{26}~Ly$\alpha$) could be partially filled with external scattered photons, apparently disguising its true line strength or EW. This can explain the fact that none of the pronounced X-ray winds in the major BH transients exhibit saturation in their absorption features even from high-polar angle sources \citep[e.g.,][]{Miller2015a}. Although it is beyond the scope of the current investigations in this work, these unique cases can be further exploited by our simulations in an effort to fit observed wind spectra of various BH XRBs \citep[e.g.][]{Miller2015a} in future work.
\subsection{P-Cygni Features}\label{Section0404}
We demonstrated explicitly that a total spectrum can exhibit, depending on the wind density, a P-Cygni profile, especially in the case of a high polar angle (\hyperref[Figure0008]{Figure~8}). In high/soft state spectra of a handful of BH transients that exhibit X-ray winds, strong absorption lines of \ion{Fe}{25} and \ion{Fe}{26} are often accompanied by a putative emission feature \citep[e.g.,][]{Miller2015a}; e.g., relatively prominent in GRO~J1655-40 and GRS~1915+105 while rather tentative in 4U~1630-472 and H~1743-322. Detailed spectral analysis of the P-Cygni profile often favors an equatorial wind geometry \citep[e.g.,][]{Miller2015a}, which is at least qualitatively consistent with our models considered here along with other alternative wind scenarios \citep[e.g.][]{Dorodnitsyn2009, Dorodnitsyn2010}. 

A quantitative comparison with observations would further require exploratory spectral simulations over a wide range of wind model parameters, which is beyond the scope of the current objective. However, targeting a specific source (e.g., GRO~J1655-40), new observations with XRISM Resolve at microcalorimeter resolution ($\sim 5$ eV) could deliver state-of-the-art spectra with a much more precise measurement of the P-Cygni structure. Spectral simulations with MONACO should be able to better constrain the underlying wind property and its geometry.\clearpage
\section{Conclusion}\label{Section0500}
We performed three-dimensional X-ray radiative transfer calculations based on self-similar solutions of magnetically driven disk winds to evaluate the effect of scattering. The key findings are as follows.

\begin{enumerate}
\item The transmitted component of the X-ray spectrum strongly depends on the polar angle. If the polar angle is $50\degr$, we detected the blueshifted absorption lines of \ion{Fe}{24} ($6.663 \ \mathrm{keV}$), \ion{Fe}{25} ($6.703 \ \mathrm{keV}$), \ion{Fe}{26} Ly$\alpha$2 ($6.959 \ \mathrm{keV}$) and \ion{Fe}{26} Ly$\alpha$1 ($6.980 \ \mathrm{keV}$). These correspond to the speed of $90 \ \mathrm{km} \ \mathrm{s}^{-1}$, $130 \ \mathrm{km} \ \mathrm{s}^{-1}$, and $300 \ \mathrm{km} \ \mathrm{s}^{-1}$, respectively. The profile of the absorption lines differs for each absorption line. The profile of the \ion{Fe}{25} absorption line is an almost symmetric Gaussian-like shape, while that of \ion{Fe}{26} is asymmetric and has a tail on the blue side. 

\item The total X-ray spectrum strongly depends on the angle of inclination. If the polar angle is less than $30\degr$, the total spectrum shows only emission lines. When the polar angle is greater than $70\degr$, the total spectrum consists of a series of rich absorption lines. In particular, the depths of these absorption lines are determined by the scattered component.

\item The absorption lines of \ion{Fe}{25} and \ion{Fe}{26} are generated in the region of the polar angle of $50\degr$ to $80\degr$. On the other hand, the absorption lines of \ion{Fe}{23} and \ion{Fe}{24} are produced in the region of the polar angle of $60\degr$ to $80\degr$.

\item We compared the effective EW of the \ion{Fe}{26} resonance lines with and without scattering. The EW with the scattered component is $20$--$50 \ \mathrm{eV}$ smaller than that of the transmitted component alone. This is because the effective depths of the \ion{Fe}{26} absorption lines are determined by the degree of scattered component in the case of high angle of inclination.
\end{enumerate}
\begin{acknowledgments}
The authors thank the anonymous referee who provided useful and detailed comments. A.T. and the present research are partly supported by the Kagoshima University postdoctoral research program (KU-DREAM). This work is also supported by the Grant-in-Aid for Early Career Scientists grant Nos. 23K13147 (A.T.) and 24K17104 (S.O.). K.F. acknowledges support in part from NASA XGS grant (80NSSC23K1021) and the Department of Physics and Astronomy of the College of Science and Mathematics at James Madison University. F.T. and M.L. acknowledge funding from the European Union - Next Generation EU, PRIN/MUR 2022 (2022K9N5B4). Numerical computations were performed on Cray XD2000 at the Center for Computational Astrophysics, National Astronomical Observatory of Japan.
\end{acknowledgments}\clearpage
\bibliography{tanimoto}{}
\bibliographystyle{aasjournal}
\end{document}